\documentclass[jcs]{iosart2x}

\usepackage{pifont}
\usepackage{amsmath,amssymb,amsfonts}
\usepackage{graphicx}
\usepackage{textcomp}
\usepackage{xcolor}  
\usepackage{algorithm,algcompatible}
\usepackage{listings}
\usepackage{makecell}
\usepackage{url}
\usepackage{multirow}

\definecolor{codegreen}{rgb}{0,0.6,0}
\definecolor{codegray}{rgb}{0.5,0.5,0.5}
\definecolor{codepurple}{rgb}{0.58,0,0.82}
\definecolor{backcolour}{rgb}{0.95,0.95,0.92}
\lstdefinestyle{mystyle}{
  backgroundcolor=\color{backcolour}, commentstyle=\color{codegreen},
  keywordstyle=\color{magenta},
  numberstyle=\tiny\color{codegray},
  stringstyle=\color{codepurple},
  basicstyle=\ttfamily\footnotesize,
  breakatwhitespace=false,         
  breaklines=true,   
  captionpos=b,      
  keepspaces=true,   
  numbers=left,      
  numbersep=4pt,    
  showspaces=false,  
  showstringspaces=false,
  showtabs=false,    
  tabsize=1,
  escapeinside={(*@}{@*)} 
}

\algnewcommand\algorithmicreturn{\textbf{return}}
\algnewcommand\RETURN{\algorithmicreturn}
\algnewcommand\algorithmicprocedure{\textbf{procedure}}
\algnewcommand\PROCEDURE{\item[\algorithmicprocedure]}%
\algnewcommand\algorithmicendprocedure{\textbf{end procedure}}
\algnewcommand\ENDPROCEDURE{\item[\algorithmicendprocedure]}%
\algnewcommand{\algvar}[1]{{\text{\ttfamily\detokenize{#1}}}}
\algnewcommand{\algarg}[1]{{\text{\ttfamily\itshape\detokenize{#1}}}}
\algnewcommand{\algproc}[1]{{\text{\ttfamily\detokenize{#1}}}}
\algnewcommand{\algassign}{\leftarrow}

\newcommand{\etal}{\textit{et al.}}
\newcommand{\ie}{\textit{i.}\textit{e.}}

\def\BibTeX{{\rm B\kern-.05em{\sc i\kern-.025em b}\kern-.wanq08em
    T\kern-.1667em\lower.7ex\hbox{E}\kern-.125emX}}

\pubyear{0000}
\volume{0}
\firstpage{1}
\lastpage{1}

\begin{document}
\numberlinesfalse

\begin{frontmatter}
\title{SQLaser: Detecting DBMS Logic Bugs with Clause-Guided Fuzzing}
\runtitle{SQLaser: Detecting DBMS Logic Bugs with Clause-Guided Fuzzing}


\begin{aug}
\author[A,B]{\fnms{Jin} \snm{Wei}\ead[label=e1]{jwei17@fudan.edu.cn}}
\author[B,C]{\fnms{Ping} \snm{Chen}\ead[label=e2]{pchen@fudan.edu.cn}\thanks{Corresponding author. \printead{e2}.}}%
\author[E]{\fnms{Kangjie} \snm{Lu}\ead[label=e5]{}}
\author[D]{\fnms{Jun} \snm{Dai}\ead[label=e3]{}}
\author[D]{\fnms{Xiaoyan} \snm{Sun}\ead[label=e4]{}}

\address[A]{School of Computer Science, \orgname{Fudan University}, Shanghai, \cny{China}\printead[presep={\\}]{e1}}
\address[B]{Institute of BigData, \orgname{Fudan University}, Shanghai, \cny{China}\printead[presep={\\}]{e2}}
\address[C]{Purple Mountain Laboratories, Nanjing, \cny{China}}
\address[E]{University of Minnesota Twin Cities, Minneapolis, \cny{USA}}
\address[D]{Worcester Polytechnic Institute, Massachusetts, \cny{USA}}

\end{aug}

\begin{abstract}
Database Management Systems (DBMSs) are vital components in modern data-driven systems. Their complexity often leads to logic bugs, which are implementation errors within the DBMSs that can lead to incorrect query results, data exposure, unauthorized access, etc., without necessarily causing visible system failures. 
Existing detection employs two strategies: rule-based bug detection and coverage-guided fuzzing. 
In general, rule specification itself is challenging; as a result, rule-based detection is limited to specific and simple rules. Coverage-guided fuzzing blindly explores code paths or blocks, many of which are unlikely to contain logic bugs; therefore, this strategy is cost-ineffective. 
In this paper, we design SQLaser, a SQL-clause-guided fuzzer for detecting logic bugs in DBMSs.
Through a comprehensive examination of most existing logic bugs across four distinct DBMSs, excluding those causing system crashes, we have identified 35 logic bug patterns.
These patterns manifest as certain SQL clause combinations that commonly result in logic bugs, and
behind these clause combinations are a sequence of functions.
We therefore model logic bug patterns as error-prone function chains (\ie, sequences of functions). We further develop a directed fuzzer with a new path-to-path distance-calculation mechanism for effectively testing these chains and discovering additional logic bugs. 
This mechanism enables SQLaser to swiftly navigate to target sites and uncover potential bugs emerging from these paths. Our evaluation, conducted on SQLite, MySQL, PostgreSQL, and TiDB, demonstrates that SQLaser significantly accelerates bug discovery compared to other fuzzing approaches, reducing detection time by approximately 60\%. As a standalone fuzzer, SQLaser identified 22 bugs spanning 18 of the 35 logic bug patterns, outperforming contemporary fuzzers like SQLRight, which only uncovered 2 logic bugs across 2 patterns within the same testing period (\ie, 60 days) when testing SQLite. Notably, 4 of the bugs discovered by SQLaser are zero-day, all of which have been reported to and confirmed by vendors.
\end{abstract}

\end{frontmatter}


\section{Introduction} \label{Introduction}

Database Management Systems (DBMSs)~\cite{DBMS} are widely used as centralized storage and management systems for data. For instance, SQLite~\cite{SQLite} is used on around 3.5 billion smartphones worldwide. 
Given their critical role of DBMSs in modern data-driven applications, 
logic bugs~\cite{SQLRight} within DBMSs can have severe consequences. These bugs, which are due to implementation errors in the DBMSs' code and are particularly common, can cause a variety of serious issues, such as incorrect query results, exposure of sensitive data, unauthorized access, and data corruption~\cite{NoREC, TLP, PQS, SQLRight}.

Due to the non-crashing nature of most logic bugs, detecting them can be challenging, and researchers have invested significant efforts in this area.
Rigger \etal~propose oracles to detect incorrectness of SQL query results. For example, the Non-Optimizing Reference Engine Construction (NoREC) oracle~\cite{NoREC} and the Ternary Logic Partitioning (TLP) oracle~\cite{TLP} are two oracles employing the concept of differential testing~\cite{mckeeman1998differential}. They transform the original SQL query into an equivalent form, and if discrepancies arise between the results of the original query and the transformed query, potential logic bugs may be indicated. Pivoted Query Synthesis (PQS) oracle~\cite{PQS} does not employ the concept of differential testing; instead, it automatically generates queries for which they ensure fetching a specific row, called the pivot row. Failure to retrieve the pivot row indicates a potential bug in the system.
SQLancer~\cite{SQLancer,SQLancer1} that implements these oracles discovers logic bugs in various DBMSs.
However, as SQLancer generates SQL queries based on specific rules, it may constrain the exploration of broader code paths. 
Therefore, Liang \etal~introduce SQLRight~\cite{SQLRight}, aiming to generate high-quality query statements with validity-oriented mutations to uncover more logic bugs.
Specifically, SQLRight employs coverage-guided fuzzing~\cite{AFL,Honggfuzz,LibFuzzer,Serebryany,Driller,QSYM,Steelix,Angora,GREYONE} to achieve higher code coverage in code paths.

\begin{lstlisting}[language=SQL, linewidth={\linewidth}, emphstyle={\color{blue}},label={code:1},caption={An example logic bug in SQLite.}]
CREATE TABLE t0(c0 COLLATE NOCASE, c1); 
CREATE INDEX i0 ON t0(c0) WHERE c0 >= c1; 
INSERT INTO t0 VALUES('a', 'B');
SELECT * FROM t0 WHERE t0.c1 <= t0.c0; 
-- output: {}, expected: {a|B} 
\end{lstlisting}

Despite substantial efforts dedicated to detecting logic bugs, we find some insights that have not been taken into account by existing research. Specifically, we observe that many logic bugs tend to follow particular patterns, often involving a combination of the effects of multiple SQL clauses.
For instance, Code~\ref{code:1} is a logic bug in SQLite. In this example, 
The first statement creates a table \texttt{t0} with two columns, \texttt{c0} and \texttt{c1}, where the values in \texttt{c0} are case-insensitive. The second statement creates an index \texttt{i0}, which only applies to values where \texttt{c0 >= c1}. The third statement inserts two values \texttt{\{a|B}\}. Since the ASCII value of 'a' is greater than the ASCII value of 'B', it satisfies the condition of the fourth query statement. Therefore, the intention of the \texttt{SELECT} statement is to retrieve all records from the \texttt{t0} table.
However, no rows are returned. The cause of this bug is that SQLite incorrectly applies a NOCASE operation, leading to an incorrect result of the conditional expression  \texttt{t0.c1} $ \leq$ \texttt{t0.c0}. In this case, column \texttt{c0} has two properties: case independence defined by the ``NOCASE'' clause in the first statement, and the creation of a partial index \texttt{i0} on \texttt{c0} as defined by the ``INDEX i0'' clause in the second statement. The interaction between the code implementing these two properties triggers a logic bug in the code path responsible for this functionality, resulting in an inaccurate output. 
We refer to these requirements that give rise to bug manifestation as the combined effect of multiple clauses at the SQL query level. Correspondingly, at the source code level, bug manifests as the combined action of multiple functions.

However, SQLRight does not consider these insights when selecting test cases (seeds) for coverage-guided fuzzing. It assumes an equal likelihood of triggering a logic bug for each seed. As a result, a portion of the explored code is often free of bugs. Therefore, relying solely on coverage-based fuzzing proves less effective. Instead, we shift towards targeting vulnerable code paths for more fruitful results.
Compared to coverage-guided fuzzing, directed fuzzing~\cite{AFLGo,Hawkeye,Fuzzguard,Leopard,Parmesan,Lee,WindRanger} 
focuses on quickly reaching and testing predefined target sites. Building upon this concept, our aim is to leverage the SQL clause combinations 
to identify error-prone code targets, thereby efficiently identifying seeds that align with the triggering bug patterns. They are more likely to produce logic bugs. 
To achieve this goal, we need to address three key problems: 
\ding{182} How can we identify bug patterns manifested at the SQL level?
\ding{183} How can we incorporate SQL-level bug patterns as the target sites for directed fuzzing?
\ding{184} How can we design a directed fuzzing tool capable of rapidly reaching and testing predefined target sites? 

To address the problem \ding{182}, we conduct \emph{the first systematic study} on most of existing logic bugs (with the exception of those leading to system crashes) in four different DBMSs, which include 144 bugs in SQLite, MySQL, PostgreSQL, and TiDB. 
Our investigation reveals that certain SQL clauses or combinations of clauses are prone to causing logic bugs. We categorize the SQL clauses relevant to logic bug into distinct groups, such as table elements/schemas, data processing functions, conditional expressions, special keywords, and query optimization functions. This way, we gather a list of 35 essential SQL-level bug patterns (\ie, SQL clause combinations) associated with the existing logic bugs.

To address the problem \ding{183}, we dynamically
trace and analyze the code path of statements that leads to logic bugs. 
Through our analysis, we further discover that executing SQL statements related to the logic bugs usually involves executing a series of functions (\ie, a \emph{functions call chain}) that implement these SQL-level bug patterns. Therefore, we attempt to formalize the SQL-level bug patterns as specific function call chains and use them as target sites in directed fuzzing.
While dynamic analysis methods enable us to obtain the complete function call chain for the statements triggering bugs, this call chain often includes numerous functions that are unrelated to SQL-level bug patterns.
Also, using a specific call chain with an excessive number of functions as the target for fuzzing may limit the exploration capability of discovering new bugs.
Therefore, we have to generalize the call chain by extracting
the inherent buggy sequence of calls. To this end, we present a method to ``trim'' the function call chains by tracking the flow of data objects related to the SQL clauses combination. By identifying the functions that manipulate these data objects, 
we can isolate specific sections of the call chain involved in processing these objects.
Consequently, we strategically select these sections of the call chain as the target for fuzzing, allowing for more effective and focused exploration.

To address the problem \ding{184}, we introduce SQLaser, a \emph{clause-guided model} for detecting logic bugs in DBMSs. SQLaser adopts a new approach to calculate the distance between seeds and target sites, \ie, it involves determining the call chains of these seeds and computing their distances to the target call chains. Furthermore, SQLaser utilizes a distinction-based approach to prioritize seeds that are closer to the target call chains. Ultimately, SQLaser repurposes SQLRight's testing oracles~\cite{SQLRight} to detect logic bugs by generating variant forms of the original queries and evaluating their execution results. This approach allows for effective bug detection by exploring different query structures while ensuring the preservation of their intended functionality.


We evaluate SQLaser on four popular DBMS systems: SQLite~\cite{SQLite}, MySQL~\cite{MySQL}, PostgreSQL~\cite{PostgreSQL} and TiDB~\cite{TiDB}. The results show that SQLaser reproduces all 35 known logic bug patterns (\ie, SQL clause combinations) derived from the collective discoveries of existing tools, and successfully discovers 22 bugs across 18 of the 35 bug patterns. Among the 22 discovered bugs, 4 are new unique bugs that have eluded detection by other fuzzers. We reported these 4 bugs, and all of them have been confirmed by vendors. Furthermore, SQLaser significantly (by approximately 60\%) outperforms SQLRight and traditional directed fuzzing approach in terms of efficiency.

In conclusion, this paper makes the following contributions:

\begin{itemize}
\item[$\bullet$] We conduct the first systematic study of most existing logic bugs in DBMSs. It reveals that logic bugs stem from specific combinations of SQL clauses,
and behind the clauses are sequences of critical functions. 
This finding motivates us to propose \emph{clause-guided fuzzing} which enables the fuzzing to focus on 
error-prone function chains to cost-effectively find logic bugs.

\item[$\bullet$] We propose a new fuzzing-feedback mechanism that
determines the distance between seed call chains and target call chains, enabling SQLaser to prioritize seeds based on their distance to the target paths. Such a path-to-path distance
calculation differs from the traditional single-point distance 
calculation and allows \emph{path-based directed fuzzing}.

\item[$\bullet$]  We present SQLaser, a tool that harnesses directed fuzzing to efficiently navigate to the target sites and automatically generate input queries for verifying if they can trigger logic bugs. We design experiments to validate the effectiveness and performance of SQLaser. By targeting the call chains that may trigger logic bugs, we are able to discover more/new logic bugs in DBMS. Moreover, compared to coverage-guided fuzzing and traditional directed fuzzing approaches, SQLaser improves efficiency by approximately 60\%.
\end{itemize}

\section{Background}
\subsection{Logic Bugs and Threat Model}
DBMS logic bugs arise from implementation errors of DBMSs, and they can be triggered by an attacker executing normal and legitimate SQL queries. Most of these bugs do not cause program crashes, making them harder to detect.
For example, in Code 1, a logic error demonstrates how a user attempting to read a value from the DBMS receives an empty result instead. This issue affects the normal functionality and usability of the DBMS.

Furthermore, logic bugs can have more severe consequences. In Code~\ref{code:impact}, the first statement creates a table \texttt{person} with only one column \texttt{pid} of type \texttt{INT}. The second statement inserts three rows into this table with values 1, 10, and 10. The third statement queries the rows with \texttt{pid=10} and, with the \texttt{DISTINCT} keyword, expects to remove duplicate rows. Based on the table content, the result should be {10}. However, the DBMS returns two values (10) and (10) because it incorrectly ignores the \texttt{DISTINCT} keyword. If the query deliberately employs the \texttt{DISTINCT} keyword to conceal the count of matched rows for privacy reasons, this bug will expose that information. Imagine a scenario where the query is part of a system that distributes random passwords to users. If \texttt{pid} represents user IDs and each user should receive a unique password, this bug could cause multiple users (two users with \texttt{pid=10} in this case) to receive the same password. This type of logic bugs can lead to critical data leakage, similar to CVE-2012-2081~\cite{CVE-2012-2081}, which revealed private group titles to non-members. Moreover, if the leaked data includes privileged information, it can facilitate privilege escalation, allowing attackers to execute further operations, akin to the impact of CVE-2014-4987~\cite{CVE-2014-4987}, which allowed unprivileged users to view the MySQL user list.
Additionally, unique bugs discovered using our tool SQLaser, such as the MySQL ANY logic bug (Code~\ref{code:4}), and the MySQL LIKE logic bug (Code~\ref{code:5}), demonstrate that logic bugs can leak data that the DBMS should not expose to users. If this data pertains to DBMS security or privileged information, it poses a significant security risk, given the widespread use of DBMSs like SQLite and MySQL.
\begin{lstlisting}[language=SQL, linewidth={\linewidth}, emphstyle={\color{blue}},label={code:impact},caption={A logic bug in SQLite can lead to data leakage.}]
CREATE TABLE person (pid INT);
INSERT INTO person VALUES (1), (10), (10);
SELECT DISTINCT pid FROM person WHERE pid=10;
-- output: {10,10}, expected: {10}
\end{lstlisting}

\subsection{Logic Bug Testing Oracles and Differential Testing}
Different from memory errors that can lead to program crashes, most logic bugs do not cause system crashes; they typically result in incorrect outcomes. Therefore, given the characteristics of logic bugs, we require an oracle to detect SQL query result correctness. While advanced analysts can manually analyze results to identify logic bugs, this approach is impractical for large-scale test cases. To address this, researchers consider using differential testing to automatically verify the correctness of SQL query results~\cite{ghit2020sparkfuzz,lo2010framework,slutz1998massive}. 

Differential testing~\cite{mckeeman1998differential} is a technique where a single input is passed to multiple functionally equivalent systems, with the expectation that these systems should produce identical outputs. If discrepancies are found among the results produced by these systems, it indicates the presence of at least one system containing a bug. However, constructing an oracle that covers all DBMS logic bugs is challenging. Additionally, due to various languages and extensions, popular DBMSs share only a small subset of functionalities, making cross-DBMS validation incapable of detecting DBMS-specific bugs~\cite{slutz1998massive}.

Researchers, including Manuel Rigger~\etal, have made efforts in this direction. They build functionally equivalent SQL queries for a single DBMS and check if these queries could produce the same results. For instance, the Non-Optimizing Reference Engine Construction (NoREC) oracle~\cite{NoREC} translates an optimized query containing a WHERE clause into an unoptimized version with identical semantics. By comparing the results obtained from both queries, inconsistencies can indicate the presence of a logic bug. Ternary Logic Partitioning (TLP) oracle~\cite{TLP} composes several sub-queries to collectively achieve the semantics of the original query. If the composed query does not yield the same result as the original query, it suggests the possibility of a logic bug. 
SQLRight~\cite{SQLRight} also supplements their approach with some oracles. The INDEX oracle adds various CREATE INDEX clauses to the query set to test the impact of INDEX usage on query execution. The ROWID oracle inserts WITHOUT ROWID clause to observe its effect on query execution.
If the results of the queries with the inserted INDEX or WITHOUT ROWID clause differ from those without these insertions, it indicates a potential logic bug. And the LIKELY oracle checks if the original queries include the LIKELY clause. If they don't, it adds the LIKELY clause to see how it impacts query execution.

It is worth noting that the Pivoted Query Synthesis (PQS) oracles~\cite{PQS} proposed by Manuel Rigger~\etal~does not use the differential testing method; instead, it automatically generates queries for which they ensure fetching a specific row, called the pivot row. If the DBMS fails to fetch the pivot row, the likely cause is a bug in the DBMS. As of now, we have not deployed PQS oracles into SQLaser.

\subsection{Coverage-guided Fuzzing and Directed Fuzzing}
Fuzzing~\cite{fuzzing} is a software testing technique that involves providing invalid, unexpected, or random data as inputs to a computer program. The goal is to discover vulnerabilities or bugs by observing how the program reacts to these inputs~\cite{Honggfuzz,LibFuzzer,Serebryany,AFL}. Coverage-guided fuzzing is a specific type of fuzz testing that utilizes code coverage feedback from the program's execution to guide the generation of new test inputs. The key idea is to measure the code coverage achieved by the current set of inputs and use this information to prioritize the generation of inputs that explore new or uncovered code paths. This approach aims to maximize the exploration of the program's code and increase the chances of discovering hidden vulnerabilities. Coverage-based fuzz testing has been applied to test various types of programs, including but not limited to operating systems~\cite{NTFUZZ,HFL,MoonShine,Krace,FileSystems}, compilers~\cite{OneEngine,CodeAlchemist,FuzzingJavaScrip}, web browsers~\cite{ClusterFuzz,Serebryany,FREEDOM}, document readers~\cite{Favocado,Cooper}, and even smart contracts~\cite{ContractFuzzer,sfuzz,Harvey}.
Recent works also use coverage-guided fuzzing to test DBMS systems~\cite{APOLLO,LiuX,WangM,Squirrel, SQLRight}. For example, 
SQLRight~\cite{SQLRight} introduces a logic bug exploration methodology utilizing coverage-based fuzzing, identifying logic bugs in both SQLite and MySQL. 
The process begins by placing all sample queries into a queue. The fuzzer then chooses queries from this queue for mutation. If the mutated queries leads to previously unexplored execution paths, they are added back to the queue for further consideration and undergoes testing to uncover any logic bugs.

Directed fuzzing~\cite{AFLGo,Hawkeye,Fuzzguard,Leopard,Parmesan,Lee,Leopard}, on the other hand, involves providing targeted or specific inputs to the program in an attempt to explore particular code paths or functionalities. Unlike coverage-guided fuzzing, which relies on random or mutated inputs, directed fuzzing is more focused and intentional. This approach is often used when there is prior knowledge about potential areas of vulnerability or when testing specific features of the program.
In this paper, we employ directed fuzzing to deliberately detect logic bugs, rather than aimlessly testing code paths. Specifically, we target logic bug patterns, which denote paths in the source code that are highly prone to triggering logic bugs. We allocate a substantial portion of the time budget to explore these targeted paths, with the objective of uncovering more bugs.

\section{Analysis of Existing Logic Bug Patterns} \label{Analysis of Existing logic bugs}

To develop an effective DBMS fuzzer for logic bugs,
we conduct the first systematic study against existing 
logic bugs. In particular, we analyze a total of 144 existing logic bugs, including 
102 in SQLite, 15 in MySQL, 1 in PostgreSQL, and 26 in TiDB~\cite{SQLRight,SQLancer1}. This analysis covers most logic bugs, and we do not investigate logic bugs that result in system crashes because we focus on logic bugs that do not lead to observable system failures (such as crashes). 

Based on the semantics of SQL clause, we classify SQL clauses related to these logic bugs into five categories: table element/schema, data processing clause, conditional expressions, special keyword, and query optimization function, as shown in Table~\ref{tab:3}. 
Below introduces each category:

\begin{itemize}
\item[$\bullet$] \emph{Table element/schema}: This category involves elements or schemas in the created tables, such as column indexes, column values, and primary keys. 

\item[$\bullet$] \emph{Data processing functions}: This category involves data processing functions. For example, 
the CAST function is used to convert value of one data type to another compatible one. The ROUND function is to round a numerical value.

\item[$\bullet$] \emph{Conditional expressions}: These types of expressions are often used for conditional checks in queries. For instance, the expression EXISTS checks whether the query result satisfies the condition expression. It returns true if at least one row is found and false otherwise. Another example is INSERT OR FAIL, which is a feature that causes the insertion of data to fail if it violates any constraints. It ensures data integrity by preventing insertion of invalid or duplicate records. 

\item[$\bullet$] \emph{Special keywords}: In SQL queries, there are certain special keywords that are used for setting constraints, referencing specific objects, or defining virtual tables or special index structures, etc. For example, WITHOUT ROWID is used to define a table that no longer has a default ROWID column. NOCASE is used to indicate that the values in a column are case-insensitive when performing comparison operations. 

\item[$\bullet$] \emph{Query optimization functions}: These functions provide the ability to further optimize and control query execution in order to improve query efficiency and performance. However, these functions may occasionally return unexpected or incorrect values. For example, the LIKELY function is used to indicate to the query optimizer that a certain condition is likely to occur. Similarly, the UNLIKELY function is used to indicate a condition that is unlikely to occur. 

\end{itemize}

\begin{table}[]
\caption{Five categories of SQL clauses and examples of clauses in each category.}
\label{tab:3}
\scriptsize
\centering
\begin{tabular}{cc}
\hline
\bf{\emph{Type}} & \bf{\emph{SQL Clauses}} \\
\hline
Table Element/Schema & column, INDEX, partial INDEX, PRIMARY KEY, etc. \\
Data Processing Functions & CAST, ROUND, etc. \\
Conditional Expressions & EXISTS, INSERT OR FAIL, IS TRUE, etc. \\
Special Keywords & WITHOUT ROWID, NOCASE, DISTINCT, RTRIM, PRAGMA, VIEW, rtree, etc.\\
Query Optimization Functions & LIKELY, UNLIKELY, GLOB, IN-early-out, etc \\
\hline

\end{tabular}
\end{table}

\begin{table}[]
\caption{Analysis of the SQL-level bug patterns of logic bugs for SQLite, MySQL, PostgreSQL and TiDB.}
\label{tab:7}
\scriptsize
\centering
\begin{tabular}{cccc}
\hline
\bf{\emph{DBMS}}           & \bf{\emph{SQL-level Bug Pattern (SQL Clause Combination)}}    & \bf{\emph{Number}} & \bf{\emph{Bug Example}}        \\
\hline
\multirow{17}{*}{SQLite~\cite{SQLite}} & INDEX, PRIMARY KEY, WITHOUT ROWID, NOCASE                                      &1         & 1b1dd4d4~\cite{1b1dd4d4}                 \\
                         & partial INDEX, LIKELY, IS FAIL                                                 & 11         &   5351e920~\cite{5351e920}                \\
                         & WITHOUT ROWID, PRIMARY KEY DESC                                                & 2            & f65c929~\cite{f65c929}                   \\
                         & column value, CAST, LIKELY, UNLIKELY, GLOB                                     & 26           & f9c6426~\cite{f9c6426} \\
                         & column value, MIN                                                              & 3            & faaaae49~\cite{faaaae49}                   \\
                         & column value, CAST                                                             & 8            & c0c90961~\cite{c0c90961}          \\
                         & column value, ROUND                                                            & 1            & db9acef1~\cite{db9acef1}         \\
                         & CAST, EXISTS                                                                   & 3            & 16252d7~\cite{16252d7}     \\
                         & INSERT OR FAIL                                                                & 1            & 659c551d~\cite{659c551d} \\
                         & RTRIM, PRIMARY KEY, WITHOUT ROWID                                              & 1            & 86fa0087~\cite{86fa0087}                  \\
                         & DISTINCT, ORDER BY                                                             & 3            & 6ac0f822~\cite{6ac0f822}     \\
                         & PRAGMA                                                                         & 3            & ebe4845c~\cite{ebe4845c}          \\
                         & ALTER TABLE, column value                                                     & 4            & 1685610e~\cite{1685610e}                   \\
                         & rtree, (COUNT,CAST)                                                          & 7            & f898d04c~\cite{f898d04c}  \\
                         & VIEW, INDEX                                                                    & 11           & 9c8c1092~\cite{9c8c1092}     \\
                         & IN, ORDER BY                                                                   & 7            & eb40248~\cite{eb40248}   \\
                         & Others                                                                          & 10           & 54110870~\cite{54110870}                  \\ \hline
\multirow{9}{*}{MySQL~\cite{MySQL}}   & column values                                                                  & 4         & 99122~\cite{99122} \\       
                         & ANY                                                                &  1        & \cite{SQLRight} \\  
                         & LIKE ESCAPE,XOR                                                                &  1        & 95927~\cite{95927}       \\
                         & BIGINT UNSIGNED, IFNULL                                                        &  2         &      95954~\cite{95954}             \\
                         & ``\textless{}=\textgreater{}'' comparison                                        &  1            &  95908~\cite{95908}                  \\
                         & IF(FALSE)                                                                      &  1            & 95926~\cite{95926}                   \\
                         & IN operator                                                                    &  1      &   95975~\cite{95975}                \\
                         & \&, \textless{}, and AND                                                       &  1            &     95983~\cite{95983}             \\
                         & GREATEST                                                                       &  1            &    96012~\cite{95908}                \\ 
                         & others                                                                         &  2        & 95937~\cite{95937}\\
                         
                         \hline
PostgreSQL~\cite{PostgreSQL}               & PRIMARY KEY, GROUP BY                                                          &    1          &   \cite{SQLancer}\\ \hline
\multirow{7}{*}{TiDB~\cite{TiDB}}    & columns                                                                        & 11          & 15725~\cite{15725} \\
                         & CAST, IsTrue/IsFalse                                                           & 2        &   15733~\cite{15733}  \\
                         & CHAR()                                                                         & 1          &   15986~\cite{15986}  \\
                         & collation                                                                      & 2          &     15789~\cite{15789}  \\
                         & JOIN                                                                           & 6           &   15846~\cite{15846}  \\
                         & USE\_INDEX\_MERGE                                                              & 1           &  15994~\cite{15994} \\
                         & Others                                                                         & 3          &     17814~\cite{17814} \\ \hline

\end{tabular}
\end{table}

Based on these categories of SQL clauses in Table~\ref{tab:3}, we categorize the causes and bug patterns of the existing 144 logic bugs in four DBMSs. This results in a summary of 35 bug patterns, comprising 17 for SQLite, 10 for MySQL, 1 for PostgreSQL, and 7 for TiDB. Each bug pattern is characterized by a combination of SQL clauses, as illustrated in Table~\ref{tab:7}. It presents a breakdown of SQL-level bug patterns for various types of logic bugs, including the number of bugs associated with each pattern and their respective bug IDs. 
For example, the SQLite logic bug 1b1dd4d4~\cite{1b1dd4d4} is caused by the confusion of using the index of columns associated with a WITHOUT ROWID table. In this table, a column has two indexes: a PRIMARY KEY index and a regular index (INDEX), where the regular index defines the values in the column as collating NOCASE. Therefore, we define the SQL clause combination for this bug to be associated with \emph{INDEX, PRIMARY KEY, WITHOUT ROWID}, and \emph{NOCASE}. We obtain the SQL clause combination for the remaining logic bugs using a similar approach.
As for MySQL, the first SQL clause combination pertains to column values, primarily including errors induced when handling floating-point numbers, such as a query whose predicate involves floating-point numbers yielding an incorrect result (bug: 99122~\cite{99122}). Additionally, some logic bugs are caused by query optimization functions such as ANY, LIKE, etc (bug: 95927~\cite{95927})).
Incorrect results can also be caused by conditional expressions such as IFFULL, IF(FALSE), and IN (bug: 95926~\cite{95926}).
In the case of PostgreSQL, only one logic bug was found, specifically related to the PRIMARY KEY and GROUP BY clauses~\cite{SQLancer}.
For TiDB, it can be observed that certain types of columns such as special characters (like double negation, etc.) or specific data types (like floats/boolean, etc.) are more prone to triggering logic bugs in TiDB (bug: 15725~\cite{15725}). Additionally, certain SQL clauses such as conditional expressions (bug: 15733~\cite{15733}), data processing functions (bug: 15986~\cite{15986}), and special keywords (bugs: 15789~\cite{15789}, 15846~\cite{15846}, 15994~\cite{15994}, 17814~\cite{17814}) also contribute to triggering logic bugs in TiDB.

\section{SQLaser}

\begin{figure*}[ht]
\centering
\includegraphics[width=\linewidth]{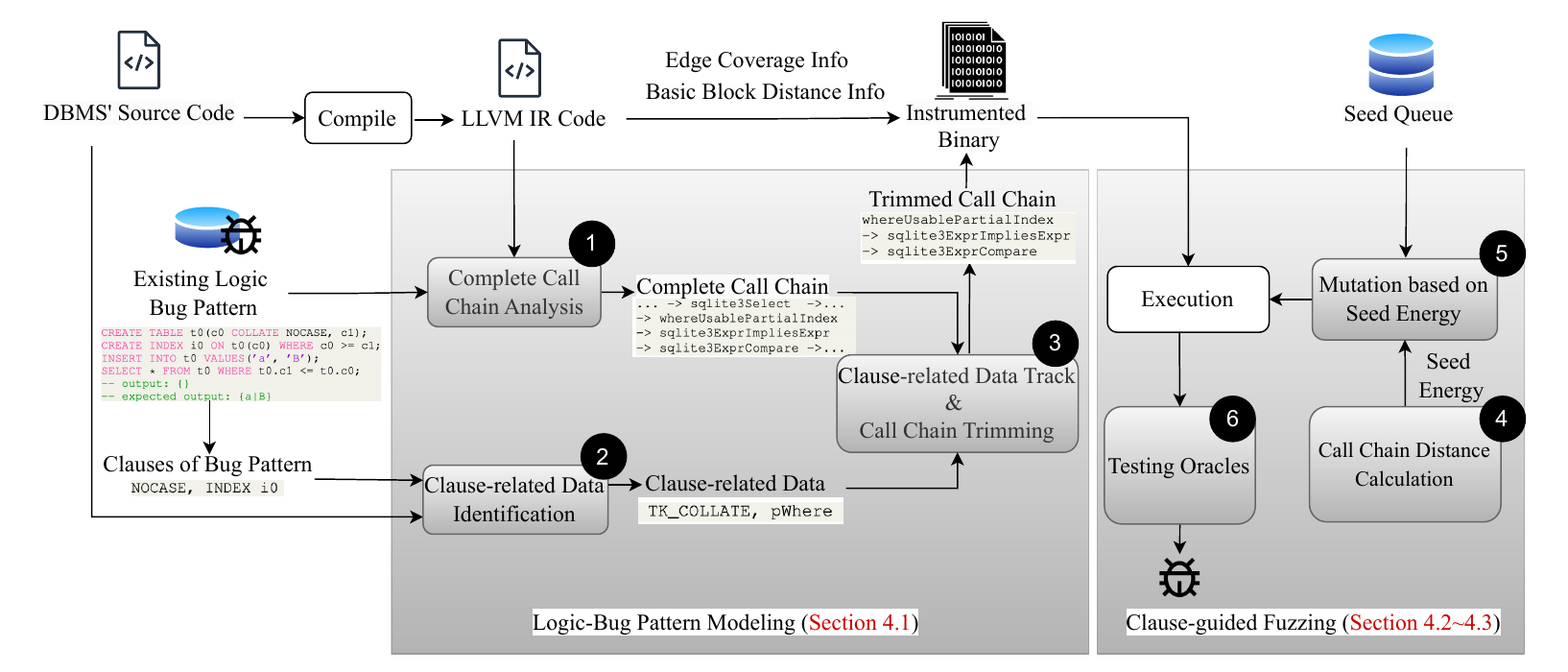}
\caption{Overview of SQLaser. \textnormal{It utilizes trimmed call chains representing SQL-level bug pattern as the target sites for the fuzzer. In the distance calculation part, we propose an algorithm to calculate the distance between seed call chains and target call chains, aiming to enable SQLaser to prioritize seeds based on their proximity to the target sites}.
}
\label{fig:1}
\end{figure*}

In this section, we present our design and implementation for SQLaser, which enables fuzzing to concentrate on error-prone function chains. These function chains are the implementation of SQL-level bug patterns, at the source code level. Specifically, our design achieves two key objectives: incorporating bug pattern information as target sites for directed fuzzing, and creating a tool with the capability to efficiently reach predefined target sites.

\textbf{Overview.}
As shown in Fig.~\ref{fig:1}, SQLaser consists of two components: 1) logic-bug pattern modeling, and 2) clause-guided fuzzing. Moreover, following established practice of existing directed fuzzing tools~\cite{AFLGo,WindRanger}, 
SQLaser instruments edge coverage information and basic block distance information into the binary code. Such information is obtained through an approach akin to WindRanger~\cite{WindRanger}, with further intricate details omitted in this paper. 

In logic-bug pattern modeling phase, SQLaser aims to instrument the call chains of bug pattern information into binary code, and these call chains manifest as target call chains during the fuzzing phase. 
Specifically, SQLaser first obtains the complete call chains executed when these bugs occur through dynamic analysis (Step \ding{182}). 
However, the complete call chain contains many functions unrelated to the bug, and excessive functions in the target call chain may limit fuzzer's exploration capability.
Thus, we identify data objects that represent the attributes of bug patterns through manual analysis, referred to as clause-related data (Step \ding{183}). Then, using automated static analysis tools, SQLaser extracts partial call chains from the complete call chain by automatically tracking clause-related data, achieving the goal of trimming the complete call chain (Step \ding{184}). The trimmed call chains are then instrumented into the binary as the target call chains for fuzzing.
This aspect will be elaborated in Section~\ref{Bug Pattern Information}.

In the clause-guided fuzzing phase, we calculate the distance between the seed call chain and the target call chain (Step \ding{185}) and determine the seed's energy based on this distance, which in turn prioritizes seeds for mutation (Step \ding{186}). After mutating the seed, SQLaser executes seeds and analyzes the results by testing oracles to determine if any logic bugs are triggered (Step \ding{187}). Finally, the mutated seeds are kept as new seeds, and are added to the seed queue for further execution. This phase will be elaborated in Section~\ref{Fuzzing} for call chain distance calculation, and in Section~\ref{Mutation and Testing Oracles} for mutation based on seed energy and testing oracles.




\textbf{Technical Challenges.}
We identify two challenges in designing SQLaser.
In the logic-bug pattern modeling phase, our challenge mainly lies in how to identify call chains of executing SQL clause combinations (\ie, SQL-level bug patterns) as the target sites for fuzzing. In the clause-guided fuzzing phase, since our target sites are function call chains, the calculation of path-to-path distance differs from traditional single-point distance calculations. We need to formulate a novel distance calculation mechanism, guided by which to determine the requisite adjustments needed for the mutation and testing oracles.

\subsection{Logic-Bug Pattern Modeling}
\label {Bug Pattern Information}
In this section, we propose our method to obtain a representative call chain for SQL clause combinations (SQL-level bug patterns). Specifically, we first analyze the complete call chain of the statements triggering bugs by employing dynamic analysis (Step \ding{182}). In this step, we write an LLVM Pass~\cite{LLVMpass} to set hooks at function entry points, and print function calling information during program execution. 
This initial analysis provides us with a comprehensive call chain. However, this call
chain, while containing information about SQL-level bug patterns, also encompasses many functions
that are unrelated to them. Moreover, excessively long target call chains can impede the discovery of new code path during the fuzzing process, as it may prioritize reproducing existing bugs over finding new ones. Therefore, it becomes imperative to trim this complete call chain to focus on portions that are more relevant to SQL-level bug patterns. 
To accomplish this, for a specific bug pattern, we manually analyze the attributes of each clause within the pattern, considering these attributes as corresponding to related data objects in the source code (the rules we follow to find these data objects during this manual analysis will be introduced in this section).
We refer to these data objects as clause-related data objects (Step \ding{183}).
Then, we trace data objects and extract relevant portions from the complete call chains that handle these data objects (Step \ding{184}). We also accomplish this step by writing an LLVM Pass~\cite{LLVMpass}. This Pass analyzes the code to find which portions of the call chain handle the clause-related data objects, which could be a function's global or local variables, parameters, return values, and more. The Pass conducts data flow analysis to trace how the clause-related data objects are passed and modified through various functions. By correlating this data flow with the call chain, the Pass identifies the specific portions of the call chain responsible for processing these data objects, filtering out irrelevant parts and focusing only on functions that directly manipulate the clause-related data.
Below, we illustrate this with an example using Code~\ref{code:1}.

\textbf{An example.} 
Using the logic bug in Code~\ref{code:1} as the example, we start by identifying that, through the analysis of this bug, the SQL clause combination of this bug is ``NOCASE'' and ``INDEX i0''. Next, we need to locate the call chain that represents this SQL clause combination.
To pinpoint the call chain responsible for this incorrect result, we initially retrieve the complete call chain of executing the \texttt{SELECT} statement by leveraging dynamic analysis. 
We use an LLVM Pass to set hooks at function entry points, and print function calling information during program execution.
The rationale behind choosing the \texttt{SELECT} statement lies in its role as the trigger for the observed incorrect output in this example.
Subsequently, we can obtain a complete call chain for executing this statement: \emph{…→ sqlite3Select → sqlite3WhereBegin → whereLoopAddAll → whereLoopAddBtree → whereUsablePartialIndex → sqlite3ExprImpliesExpr → sqlite3ExprCompare → …}.
Complete call chain illuminates the inner workings of the SQL clause combination, such as \emph{whereUsablePartialIndex}, which is a function handling partial index processing, and \emph{sqlite3ExprImpliesExpr}, defining how to compare two values in the ``NOCASE'' scenario. However, we observe that a complete call chain often involves numerous other functions.
If the number of functions in the target call chain is excessive, it might constrain the exploration capability of fuzzing. This limitation could result in a bias favoring reproducing existing logic bugs rather than discovering new ones. Consequently, we consider using a complete call chain as the fuzzing target to be inappropriate.
Therefore, we identify the relevant data objects for the ``NOCASE'' and ``INDEX i0'' clauses, \ie, the variables \texttt{TK\_COLLATE} and \texttt{pWhere}. By tracing these two variables at IR-code level, we find that only the functions $whereUsablePartialIndex$, $sqlite3ExprImpliesExpr$, and $sqlite3ExprCompare$ manipulate these tracked variables, while the other functions in the complete call chain do not. Thus, we conclude that the trimmed call chain consists of: \emph{whereUsablePartialIndex → sqlite3ExprImpliesExpr → sqlite3ExprCompare}.

\textbf{Manually identify clause-related data objects.}
Since the complete call chain contains many functions unrelated to the clauses of bug patterns, we need to manually analyze the source code to identify the bug-related functions. Fortunately, we found that the attributes of these clauses are often represented by certain variables in the program. Therefore, in Step \ding{183}, we identify data objects representing the attributes of clauses using the following rules:
\begin{itemize}
\item[$\bullet$] \emph{Table element/schema}: We identify the variables representing these elements or schemas. Like INDEX, we identify the variable \texttt{pWhere} which references this schema. Similarly, for PRIMARY KEY, we identify the variable \texttt{pPk} which points to the primary key. In the case of value in column, we identify the variable \texttt{pTab-\textgreater{}aCol} which represents the column.

\item[$\bullet$] \emph{Data processing functions}: We identify the parameters or return values of these functions. 
In the case of the CAST function, we identify the variables such as \texttt{TK\_CAST} (a flag indicating whether the CAST function is executed), \texttt{ppVal} (representing the variable involved in the conversion), and \texttt{affinity} (indicating the type of conversion).
Similarly, for the ROUND function, we identify the variable \texttt{context}, which represents the result of the ROUND operation.

\item[$\bullet$] \emph{Conditional expressions}: We identify the flags or variables that determine the conditional expressions values. As an example, for the EXISTS expression, we identify the variable \texttt{pExpr-\textgreater{}op}, which is the operator of the current expression, and check if this variable is assigned the flag \texttt{TK\_EXISTS}. Similarly, for the IS TRUE expression, we identify \texttt{pExpr-\textgreater{}op}, which in this case would be assigned the flag \texttt{TK\_TRUTH}, \ie, \texttt{pExpr-\textgreater{}op} = \texttt{TK\_TRUTH}.

\item[$\bullet$] \emph{Special keywords}: We identify the flags or variables that are used to implement the functionality of certain keywords. For instance, 
the flag \texttt{TK\_COLLATE}, \texttt{TF\_WithoutRowid}, and \texttt{EP\_Distinct} respectively indicate the usage of the NOCASE, WITHOUT ROWID, and DISTINCT keywords.
In cases involving queries with the VIEW keyword, we identify the variable \texttt{pSelTab}, which represents a VIEW used to retrieve the result set. 

\item[$\bullet$] \emph{Query optimization functions}: We identify the flags or variables associated with the implementation of optimization functions. For instance, for the LIKELY function, we identify the flag \texttt{SQLITE\_FUNC\_LIKE}, which indicates the use of the LIKELY optimization function in the query. Similarly, for the UNLIKELY and GLOB optimization functions, we can identify the flags \texttt{EP\_Unlikely} and \texttt{SQLITE\_INDEX\_CONSTRAINT\_GLOB}, respectively.
\end{itemize}

\subsection{Call Chain Distance Calculation}
\label{Fuzzing}

Our fuzzer tries to reach the target sites, \ie, function call chains. Different from traditional single-point distance calculation, here we need to define a novel distance calculation mechanism that supports path-to-path distances. 
If we consider the trimmed call chain $f_1$ $\rightarrow$ $f_2$ $\rightarrow$ ... $\rightarrow$ $f_m$ representing a SQL-level bug pattern, we can represent it using an \emph{m}-tuple model, as shown in Equation~\ref{eq:1}:
\begin{equation}
    \label{eq:1}
    targetCallChain = trimCallChain = (f_1,f_2,...,f_m)
\end{equation}

In Equation~\ref{eq:1}, $f_1$ direct calls $f_2$, $f_2$ direct calls $f_3$, ..., and $f_{m-1}$ direct calls $f_m$. 
Assuming $f_i$ direct calls $f_j$, \ie, $f_i\rightarrow f_j$, we use $ctl_{f_if_j}$ to express the control relationship of $f_i$ and $f_j$. In our model, $ctl_{f_if_j}$ is represented as a 2-tuple model:
\begin{equation}
    \label{eq:2}
    ctl_{f_if_j}=(f_i, BB_{f_i\rightarrow f_j})
\end{equation}
 
From Equation~\ref{eq:2}, the $ctl_{f_if_j}$ contains two tuples, $f_i$ and $BB_{f_i\rightarrow f_j}$. $BB_{f_i\rightarrow f_j}$ is the basic block where $f_i$ calls $f_j$ in LLVM IR code. Thus, the control relationship of this trimmed call chain is:
\begin{equation}
    \label{eq:3}
    bugCtl=(ctl_{f_1f_2}, ctl_{f_2f_3},...,ctl_{f_{m-1}f_m})
\end{equation}

In SQLaser, $bugCtl$ is the target control relationship, \ie, 
\begin{equation}
    \label{eq:5}
    targetCtl=bugCtl
\end{equation}

For a given seed, SQLaser can automatically acquire the call chain during the process of executing that seed.
If a seed's call chain involves $k$ functions, namely, $g_1, g_2,..., g_k$, just like the Equation~\ref{eq:3}, the control relationship of seed call chain is:



\begin{equation}
    \label{eq:4}
    seedCtl=(ctl_{g_1g_2}, ctl_{g_2g_3},...,ctl_{g_{k-1}g_k})
\end{equation}

And our aim is to calculate the distance between $seedCtl$ and $targetCtl$. We define the seed basic block set ($SBB$) and the target basic block set ($TBB$), respectively:
\begin{equation}
    \label{eq:8}
    SBB=\{BB_{g_1 \rightarrow g_2},BB_{g_2 \rightarrow g_3},..., BB_{g_{k-1} \rightarrow g_k}\}
\end{equation}

\begin{equation}
    \label{eq:9}
    TBB=\{BB_{f_1 \rightarrow f_2},BB_{f_2 \rightarrow f_3},..., BB_{f_{m-1} \rightarrow f_m}\}
\end{equation}

Following the idea of directed fuzzing research~\cite{AFLGo,WindRanger}, we use the distance between the basic block in $SBB$ and the basic block in $TBB$ to represent the distance between the seed call chain and the target call chain, \ie,

\begin{equation}
    \label{eq:10}
    d_{CallChain}=d_{BB}=\sum_{b\in SBB} {d(b, TBB)}
\end{equation}
where 
\begin{equation}
    \label{eq:11}
    d(b, TBB)=(\sum_{tb\in TBB} {d(b, tb)}^{-1})^{-1}
\end{equation}

In Equation~\ref{eq:11}, we use the harmonic mean to calculate the $d(b, TBB)$, and $d(b, tb)$ is the distance of basic block $b$ to $tb$. When computing the distance between two basic blocks, we employ a loose calculation approach. If the two basic blocks are within the same function, we consider their distance as $undefined$. Otherwise, the distance between the two basic blocks is equivalent to the distance of their respective functions. We express this calculation method in Equation~\ref{eq:12}.
\begin{equation}
    \label{eq:12}
    d(b, tb)=    
    \begin{cases}
    & undefined, ~\text{if}~b ~\text{and} ~ tb~\text{are in same function}, \\
    & d_f(d,tb), ~\text{otherwise}.
    \end{cases}
\end{equation}

In Equation~\ref{eq:12}, $d_f(b, tb)$ represents the distance between the function where basic block $b$ is located and the function where $tb$ is located. We denote the functions where the two basic blocks are located as $f(b)$ and $f(tb)$.
If $f(b)$ and $f(tb)$ are unreachable, the distance between them is undefined. Otherwise, it is defined as the shortest distance among all paths connecting the two functions.
This calculation is expressed in Equation~\ref{eq:13}, and $\min~d(f(b),f(tb))$ represents the shortest distance from function $f(b)$ to function $f(tb)$. 
This distance corresponds to the number of edges along the shortest path between these two functions within the function call graph.
\begin{equation}
    \label{eq:13}
    d_f(d,tb)=    
    \begin{cases}
    & undefined, ~\text{if}~$f(b)$~\text{and}~$f(tb)$ ~\text{are unreachable}, \\
    & \min d(f(b),f(tb)), ~\text{otherwise}.
    \end{cases}
\end{equation}

It is worth mentioning that due to the analysis of multiple logic bugs, we instrument multiple $targetCallChain$s. To decide which $targetCallChain$ to use for calculating the distance with the current seed, we first compute the text similarity between the $seedCallChain$ and different $targetCallChain$s. Specifically, we determine how many functions are common between the $seedCallChain$ and one $targetCallChain$, considering this counts as the text similarity between the two call chains. Finally, we follow Equation~\ref{eq:10} to calculate the distance between the $seedCallChain$ and the $targetCallChain$ with the highest textual similarity. 
In cases where multiple $targetCallChain$s exhibit the same textual similarity to the seed call chain, we select the one with the shortest distance as the distance metric between that $seedCallChain$ the $targetCallChain$.

The above seed distance calculation method details the process of determining the distance between a seed and the target call chain. This method is applied to the seed's mutation strategy, where seeds closer to the target call chain are given more mutation opportunities, as described in Section~\ref{Mutation}.

\subsection{Mutation based on Seed Energy and Testing Oracles}
\label{Mutation and Testing Oracles}

In this section, we elaborate on our mutation approach to prioritizing seeds with shorter distances, as well as the deployment of four testing oracles utilizing the differential testing method.

\subsubsection{Mutation based on Seed Energy} \label{Mutation}
In the mutation phase, SQLaser initially adopts the validity-oriented query generation method proposed by SQLRight~\cite{SQLRight}, which can effectively avoid generating queries that result in syntax or semantic errors. These generated queries serve as seeds for fuzzing, and SQLaser assigns energy to each seed. A higher energy value implies that the seed has a greater chance of undergoing substantial mutations. Specifically, SQLaser utilizes Equation~\ref{eq:10} to calculate the distance between the seed and the target. A shorter distance results in a higher energy allocation for the seed, as expressed in Equation~\ref{eq:16}. 

\begin{equation}
    \label{eq:16}
    energy={d_{CallChain}}^{-1}
\end{equation}

\subsubsection{Testing Oracles} We utilize five testing oracles to analyze the outcomes of seed execution, similar to SQLRight~\cite{SQLRight}. These oracles can transform the original queries into variant forms with different syntactic structures but equivalent functionality. By comparing the execution results of the original queries and their variants, we can detect potential logic bugs. The five oracles include: 1) the NoREC oracle, which rewrites an optimized query containing a WHERE clause to a SELECT query statement without the WHERE optimization; 2) the TLP oracle, which partitions a given query into multiple equivalent queries, and their results can be combined to obtain the same results as the original query; 3) the INDEX oracle, which introduces various CREATE INDEX clauses into the given query set, testing the impact of INDEX usage on the query execution; and 4) the ROWID oracle, which examines whether the original queries contain the WITHOUT ROWID clause. If it is not present, the WITHOUT ROWID clause is inserted to observe its effect on the query execution. 5) the LIKELY oracle checks if the original queries include the LIKELY clause. If they don't, it adds the LIKELY clause to see how it impacts query execution.

\section{Implementation}
In this section, we introduce the methods of deploying SQLaser, covering two key aspects: the instrumentation phase and the fuzzing phase.

\textbf{\emph{Instrumentation phase.}}
We instrument four types of code snippets at compilation time.
1) In order to collect code coverage during the execution process of the fuzzer, we instrument edge coverage information of tested programs. 
2) To calculate the distance between seed and target call chains, we instrument basic blocks distance information.
The above instrumentation methods are utilized similarly to Windranger~\cite{WindRanger}'s approaches.
3) To obtain the seed call chain during execution, we set hooks at every function entry points during the instrumentation phase.
4) Additionally, we instrument various logic bug patterns, \ie, corresponding call chains of SQL clause combinations. Specifically, we insert a unique identifier, which is used to distinguish different call chains, at the entry point of each function involved in a particular call chain. These identifiers can be collected and analyzed at runtime to determine if the execution can construct the target call chains. 

\textbf{\emph{Fuzzing phase.}}
In the fuzzing phase, we modify the code of WindRanger~\cite{WindRanger} to align with our design.
Specifically, we change the method for distance calculation and mutation in WindRanger, encompassing approximately \emph{10K} lines of code. Additionally, we also implement testing oracles by changing the execution analysis of WindRanger. This portion of modified code comprises around \emph{3K} lines.

\section{Evaluation}
In this section, we evaluate the following questions:
\begin{itemize}
    \item[$\bullet$] How effective is SQLaser in discovering bugs in real-world programs? Can any new unique bugs be identified? (Section~\ref{Finding Bugs in DBMSs}) 
    \item[$\bullet$] How does SQLaser compare to state-of-the-art fuzzing approaches in terms of relevant metrics, such as performance, code coverage and seed distance distribution? (Section~\ref{Comparison with Existing Tools}) 
    \item[$\bullet$] What's the effect of trimming? When the target is set to complete call chains or trimmed call chains, what are the respective numbers of false positives and false negatives generated by SQLaser?  (Section~\ref{Trimmed Complete})
\end{itemize}

\textbf{Experimental Environment}. We conduct our experiments on a machine powered by an Intel(R) Xeon(R) Gold 5118 CPU @ 2.30GHz with 16 cores. The experiments are performed on an Ubuntu 20.04.5 LTS operating system. 

\textbf{Testing DBMSs}. To compare with SQLRight, we chose to test the same versions and types of DBMSs as tested by SQLRight. Specifically, we test SQLite of version 3.28, MySQL of version 8.0.27, PostgreSQL of version 13.0. Additionally, we also conduct tests on TiDB of version 3.0.12 and SQLite of version 3.41. These two versions of the DBMS were not tested by SQLRight. 

\textbf{Experimental Settings}. In our experiment, we compare SQLaser, SQLRight, and Windranger when testing the above DBMSs, as SQLRight and Windranger represent an instance
of coverage-guided fuzzer and single-function-target directed fuzzer, respectively.
We test SQLite for 60 days, while the other DBMSs are tested for 30 days each. 
It is worth noting that both SQLRight and SQLaser are capable of detecting DBMS logic bugs. We compare the performance of the two tools in discovering logic bugs by comparing the number of logic bugs found by each tool within the same period.
However, Windranger cannot detect logic bugs. Therefore, as a directed fuzzing tool, we set Windranger's target to individual functions where the patch of an existing logic bug is located. This allows us to compare the time taken to reach the target call chain between Windranger and SQLaser.
In addition, we also compare the code coverage and seed distance distribution of SQLaser, SQLRight, and Windranger, as shown in Section~\ref{Comparison with Existing Tools}.

\subsection{Finding Bugs in DBMSs} \label{Finding Bugs in DBMSs}
In Table~\ref{tab:6}, we summarize the bugs discovered by SQLaser. As described in Section~\ref{Analysis of Existing logic bugs}, across the four tested DBMSs, 35 bug patterns are derived from the collective findings of existing tools, comprising 17 for SQLite, 10 for MySQL, 1 for PostgreSQL, and 7 for TiDB. Remarkably, as a standalone fuzzer, we successfully reproduced all 35 bug patterns and found 22 bugs in 18 of these patterns, with SQLite accounting for 10 patterns, and MySQL and TiDB each finding 4 patterns. Among the 22 discovered bugs, we confirmed that 18 are known bugs and 4 are new zero-day bugs, consisting of 2 in SQLite and 2 in MySQL. The 2 unique new SQLite bugs are effective in both versions 3.28 and 3.41. We have responsibly reported these unique bugs, and all of them have been confirmed by vendors.

\begin{table}[]
\caption{Logic bugs found by SQLaser.}
\label{tab:6}
\scriptsize
\centering
\begin{tabular}{ccccc}
\hline
\bf{\emph{DBMS}} & \bf{\emph{SQL Clause Combination (SQL-level Bug Pattern)}} & \bf{\emph{Reproduce?}} & \bf{\emph{Bug Found?}} & \bf{\emph{Unique Bug}}? \\ \hline
\multirow{17}{*}{SQLite~\cite{SQLite}} & INDEX, PRIMARY KEY, WITHOUT ROWID, NOCASE & \checkmark                       &              &                   \\
                         & partial INDEX, LIKELY, IS FAIL  & \checkmark          &  \checkmark      &         \\
                         & WITHOUT ROWID, PRIMARY KEY DESC & \checkmark                      &              &                   \\
                         & column value, CAST, LIKELY, UNLIKELY, GLOB & \checkmark                     & \checkmark          &                   \\
                         & column value, MIN  & \checkmark                      &   \checkmark  &                   \\
                         & column value, CAST & \checkmark                    & \checkmark          &                   \\
                         & column value, ROUND & \checkmark                      & \checkmark          &                   \\
                         & CAST, EXISTS  & \checkmark                  &              &                   \\
                         & INSERT OR FAIL & \checkmark                  &              &                   \\
                         & RTRIM, PRIMARY KEY, WITHOUT ROWID & \checkmark                &              &                   \\
                         & DISTINCT, ORDER BY  & \checkmark    & \checkmark          &                   \\
                         & PRAGMA & \checkmark        & \checkmark          &                   \\
                         & ALTER TABLE, column value  & \checkmark                        &              &                   \\
                         & rtree, (COUNT,CAST)  & \checkmark           & \checkmark          & \checkmark (Code~\ref{code:2})     \\
                         & VIEW, INDEX & \checkmark         & \checkmark          & \checkmark(Code~\ref{code:3})      \\
                         & IN, ORDER BY & \checkmark          & \checkmark          &                   \\
                         & Others & \checkmark                 &              &                   \\ \hline
\multirow{9}{*}{MySQL~\cite{MySQL}}   & column values & \checkmark             & \checkmark          &                   \\
                         & ANY & \checkmark                                  & \checkmark          & \checkmark (Code~\ref{code:4})      \\
                         & LIKE ESCAPE,XOR & \checkmark                                      & \checkmark          & \checkmark (Code~\ref{code:5})       \\
                         & BIGINT UNSIGNED, IFNULL& \checkmark            &              &                   \\
                         & ``\textless{}=\textgreater{}'' comparison & \checkmark                                       &              &                   \\
                         & IF(FALSE)  & \checkmark                                                                    &              &                   \\
                         & IN operator  & \checkmark                                       & \checkmark          &                   \\
                         & \&, \textless{}, and AND  & \checkmark                                                     &              &                   \\
                         & GREATEST   & \checkmark                                                                    &              &                   \\ 
                         & others  & \checkmark                                                                     &              &  
                         \\\hline
PostgreSQL~\cite{PostgreSQL}               & PRIMARY KEY, GROUP BY & \checkmark                       &              &                   \\ \hline
\multirow{7}{*}{TiDB~\cite{TiDB}}    & columns   & \checkmark                               & \checkmark          &                   \\
                         & CAST, IsTrue/IsFalse & \checkmark                                & \checkmark          &                   \\
                         & CHAR()  & \checkmark                                        & \checkmark          &                   \\
                         & collation & \checkmark                                     & \checkmark          &                   \\
                         & JOIN     & \checkmark                                                 &              &                   \\
                         & USE\_INDEX\_MERGE   & \checkmark                                                           &              &                   \\
                         & Others & \checkmark                                                                        &              &                   \\ \hline
Total                    & 35                                   &35                                         & 18           & 4     \\
\hline
\end{tabular}
\end{table}

Here, we present 4 unique bugs discovered by SQLaser. 

\textbf{SQLite rtree module bug}. Code~\ref{code:2} displays the first bug we discovered using SQLaser. The \emph{rtree} module allows the creation and manipulation of spatial indexes in SQLite. When we insert int64 data into the \emph{rtree} table (Line 2), the query in the Line 3 returns the correct result. However, the expression (in line 5) with the same semantics as Line 3 returns inconsistent results. This is because the default behavior of the \emph{rtree} module converts data into 32-bit floating-point values, which can lead to unexpected behavior due to precision loss. Besides, in this case, the expression $v0.c3 = 9223372036854775807$ may not be computed with the same value in the two SELECT statements at Line 3 and Line 5.

\begin{lstlisting}[language=SQL, emphstyle={\color{blue}}, caption={SQLite rtree module bug.},label={code:2}]
CREATE VIRTUAL TABLE v0 USING rtree ( c1, c2, c3 );
INSERT INTO v0 ( c1, c2, c3 ) VALUES ( - 0, 9223372036854775807, 9223372036854775807 ), ( 9223372036854775807, 0, 9223372036854775807 );
SELECT COUNT ( * ) FROM v0 WHERE v0.c3 = 9223372036854775807;
--output:{2}
SELECT TOTAL ( ( CAST ( v0.c3 = 9223372036854775807 AS BOOL ) ) != 0 ) FROM v0 ;
--output:{0.0}, excepted:{2.0}
\end{lstlisting}

\textbf{SQLite MIN function bug}. In Code~\ref{code:3}, the view \texttt{v5} is composed of the minimum \texttt{c1} value and the corresponding \texttt{c2} column from rows in table \texttt{v0} where the \texttt{c1} column is empty (Line 4). When querying the view \texttt{v5} in the Line 6, it returns \texttt{\{NULL|1\}}. However, when querying \texttt{v5} in the Line 8, the expression \texttt{v5.c2} returns \texttt{FALSE}, and this query returns empty. 
This might be because the value of the \texttt{c1} column is \texttt{NULL}, and due to the presence of multiple rows in \texttt{v0}, the \texttt{MIN(c1)} function in each query may not return a consistent row. Instead, it could randomly select the \texttt{c1} column value from the \texttt{\{NULL|NULL\}} rows.

\begin{lstlisting}[language=SQL, emphstyle={\color{blue}}, caption={SQLite MIN function bug.},label={code:3}]
CREATE TABLE v0 ( c1 INT, c2 INT );
INSERT INTO v0 VALUES ( NULL, 1 );
CREATE INDEX i4 ON v0 ( c1 );
CREATE VIEW v5 AS SELECT MIN ( c1 ), c2 FROM v0  WHERE c1 ISNULL;
INSERT INTO v0 VALUES(NULL,NULL);
SELECT * FROM v5;
--output:{|1}
SELECT * FROM v5 WHERE v5.c2;
--output:{}, expected:{|1}
\end{lstlisting}

\textbf{MySQL ANY logic Bug}. MySQL provides the ANY keyword for comparing a value with a set of values. For Code~\ref{code:4}, the query in Line 3 returns a correct result, as the value of WHERE expression is \texttt{FALSE}. However, the query in Line 5 returns incorrect result, which could be caused by an error in the logic of SELECT ANY calculation.
This bug reveals the number of rows that are intended to be hidden by using inequality conditions, thereby leaking table attribute values that should not be disclosed to the user.

\begin{lstlisting}[language=SQL, emphstyle={\color{blue}}, caption={MySQL ANY logic Bug.},label={code:4}]
CREATE TABLE t0(c1 INT);
INSERT INTO t0 VALUES ( '16777215' );
SELECT COUNT( * ) FROM t0 WHERE c1 < ANY ( VALUES ROW ( 0 ), ROW ( 0 ) );
--output:{0}
SELECT c1 < ANY ( VALUES ROW ( 0 ), ROW ( 0 ) ) FROM t0;
--output:{1},expected:{0}
\end{lstlisting}

\textbf{MySQL LIKE logic bug}. MySQL LIKE keyword is used for pattern matching within fields. In Code~\ref{code:5}, it can be observed that for the two rows in the \texttt{t0} table, the values of the LIKE expression in Line 4 and Line 6 are respectively 1 (TRUE) and 0 (FALSE). However, when using the LIKE expression as a condition in the WHERE clause, it indeed incorrectly retrieves rows where the LIKE expression evaluates to FALSE. This could be indicative of a logic inconsistency in the implementation of the LIKE keyword.
This bug reveals the values of table variables that are meant to be hidden by using inequality conditions, thereby exposing stored values that should not be disclosed to the user.
\begin{lstlisting}[language=SQL, emphstyle={\color{blue}}, caption={MySQL  LIKE logic bug.},label={code:5}]
CREATE TABLE t0 ( c1 INT2 ( 4 ) UNSIGNED ZEROFILL DEFAULT '0000' NOT NULL, c2 INT2 ( 4 ) UNSIGNED DEFAULT '0' NOT NULL );
INSERT INTO t0 VALUES ( 9410, 9412 );
INSERT INTO t0 VALUES ( '127.4', '127.4' );
SELECT c2 LIKE c2 ESCAPE ( SELECT COUNT( * ) FROM t0 WHERE ( 127.4 ) <> ( c1 XOR c2 >= '0' AND c2 ) ) FROM t0;
--output:{9412|1,127|0}
SELECT c2 FROM t0 WHERE c2 LIKE c2 ESCAPE ( SELECT COUNT( * ) FROM t0 WHERE ( 127.4 ) <> ( c1 XOR c2 >= '0' AND c2 ) );
--output:{9412|127}, expected:{9412}
\end{lstlisting}

\emph{\textbf{Bug features}}.
From Table~\ref{tab:6}, it is evident that both the reproduced and newly discovered logic bugs fall within the 35 bug patterns that we summarize and instrument. We determine whether a bug is a newly unique one by analyzing its root cause. After confirmation from both our team and the DBMS developers, these four new bugs indeed have distinct root causes compared to existing logic bugs. Additionally, in our testing of SQLite and TiDB, 
we find bugs in the majority of the bug patterns. However, there are still some bug patterns where we do not find bugs, possibly due to the limited duration of our fuzzing runs.
It is expected that extending the runtime will lead to the discovery of more bugs, and this is outlined as part of our future work. 
Furthermore, we note that in the case of MySQL, we find bugs in 4 out of the 10 patterns we instrument, indicating that we do not cover the majority of the bug patterns. We attribute this to the incompleteness of testing oracles supported by SQLaser, which may not be comprehensive enough to detect all types of bugs. For instance, some bugs are identified by the PQS oracle~\cite{PQS} proposed by Manuel Rigger \etal, which is not yet integrated into SQLaser (as discussed in Section~\ref{Completeness of Testing Oracles}). Additionally, out of the 15 existing bugs we summarize for MySQL, 13 are detected by the PQS oracle, explaining the comparatively lower number of bug found by SQLaser for MySQL.

\subsection{Comparison with Existing Tools} \label{Comparison with Existing Tools}
For evaluating SQLaser, we also ran SQLRight~\cite{SQLRight} and WindRanger~\cite{WindRanger}, which represent an instance of coverage-guided fuzzer and single-function-target directed fuzzer, respectively. 
In this section, we delve into a comparison of the performance, code coverage and seed distance distribution of the three fuzzers.

\subsubsection{Performance} \label{Performance}

\begin{figure}[ht]
\centering
\includegraphics[width=0.8\linewidth]{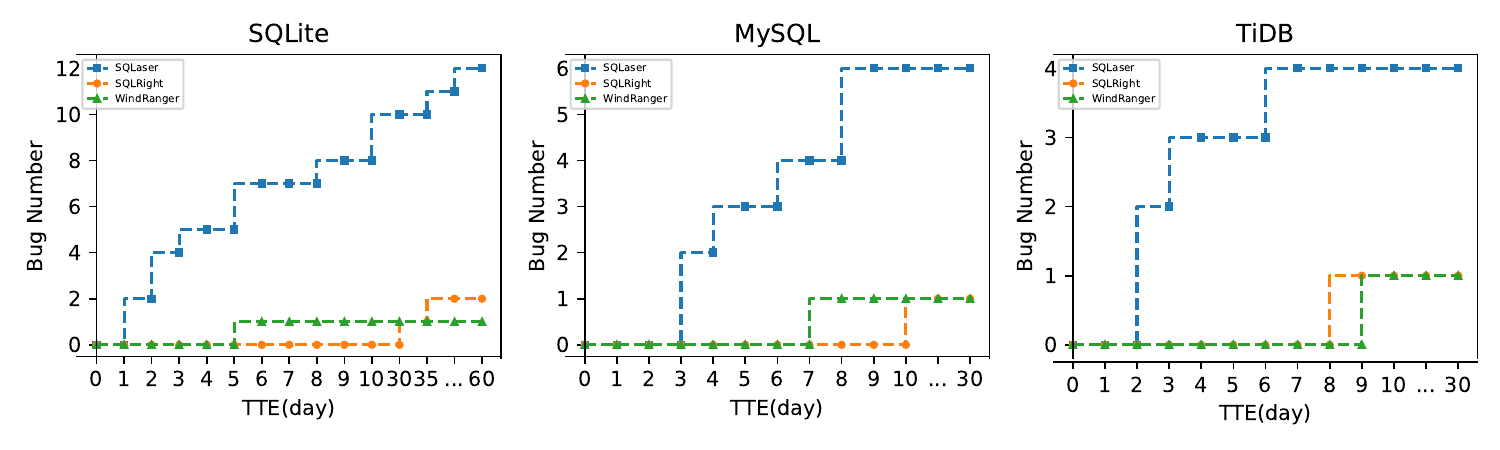}
\caption{TTE of SQLaser, SQLRight and WindRanger testing SQLite, MySQL and TiDB. Due to SQLRight and SQLaser's inability to detect bugs in PostgreSQL, we can't collect TTE statistics for PostgreSQL.}
\label{fig:2}
\end{figure}
We measure the time taken by each fuzzer to trigger bugs using the Time-to-Exposure (TTE) metric, and the results are shown in Fig.~\ref{fig:2}. As for SQLite, SQLaser completes the testing process in 60 days, during which we record the TTE for each bug triggered by SQLaser. 
We also run SQLRight for 60 days and observe that it triggers the first bug on the 30th day, and the second bug on the 35th day for SQLite. However, SQLRight does not discover any other bugs within the 60-day testing period. This indicates that SQLRight, due to its blind exploration strategy, it requires a significant amount of time to cover code paths that may lead to bug triggers.
Simultaneously, we evaluate whether WindRanger can cover the call chains of existing logic bugs. Since WindRanger does not employ a validity-oriented query generation strategy, we configure WindRanger to target the individual function where the patch for an existing logic bug is located, rather than the entire target call chain. We record the TTE for WindRanger covering the functions of target call chains. We run WindRanger for 60 days and observe that it covers the functions of one target call chain on the 5th day, but fails to cover the other target call chains.
As for other DBMSs, we also record the TTE of SQLaser, SQLRight, and WindRanger to trigger bugs during testing with MySQL and TiDB. 
Due to the fact that SQLRight and SQLaser do not find bugs in PostgreSQL, we are unable to gather TTE statistics for PostgreSQL.
Through our evaluation, we find that both SQLRight and WindRanger take longer to discover bugs or cover target call chains due to their lack of explicit targeting of the bug pattern information. In contrast, SQLaser achieves a significant decrease in bug discovery time, approximately 60\%.

\subsubsection{Code Coverage} \label{Code Coverage}
\begin{figure}[htbp]
\centering
\includegraphics[width=1\linewidth]{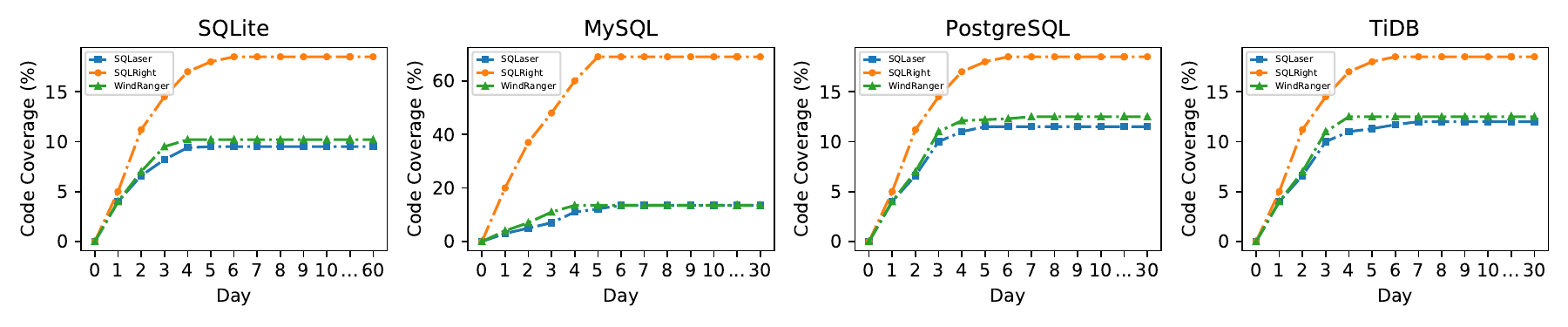}
\caption{Code coverage of SQLaser, SQLRight and WindRanger for NoREC oracle. }
\label{fig:3}
\end{figure}

\begin{figure}[htbp]
\centering
\includegraphics[width=\linewidth]{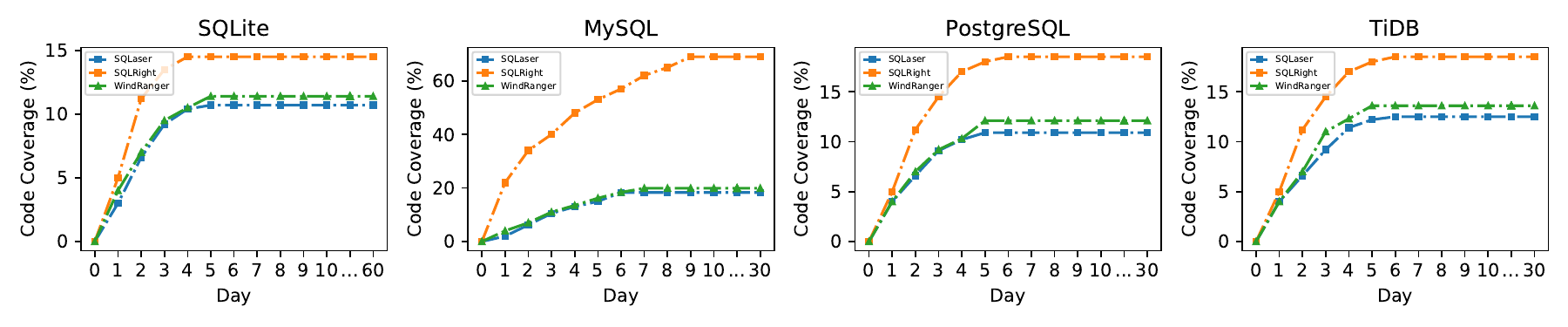}
\caption{Code coverage of SQLaser, SQLRight and WindRanger to trigger bugs for TLP oracle.}
\label{fig:4}
\end{figure}

\begin{figure}[htbp]
\centering
\includegraphics[width=0.8\linewidth]{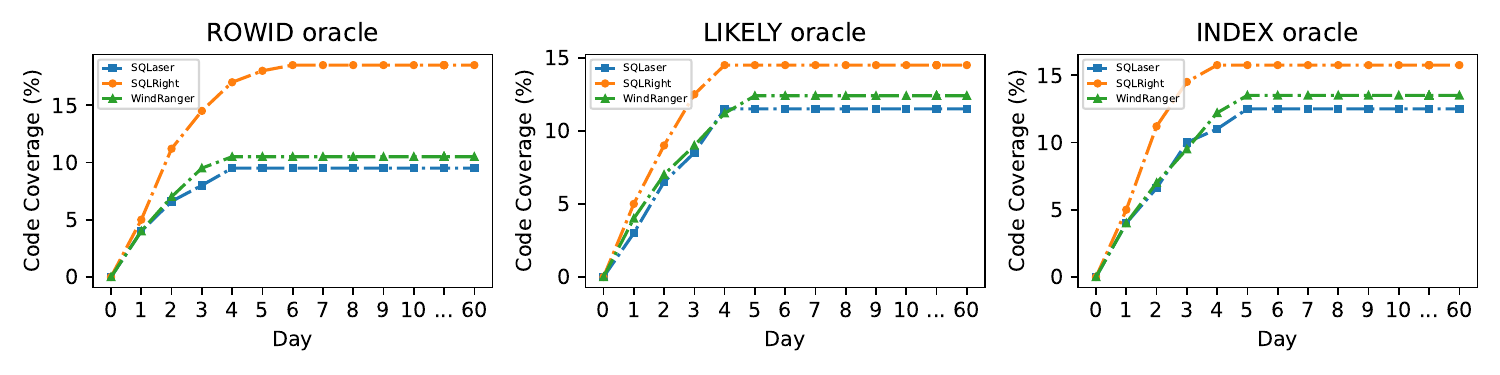}
\caption{Code coverage of SQLaser, SQLRight and WindRanger to trigger bugs in SQLite for ROWID, LIKELY and INDEX oracle. }
\label{fig:5}
\end{figure}
As depicted in Fig.~\ref{fig:3}, Fig.~\ref{fig:4} and Fig.~\ref{fig:5}, we record the code coverage of SQLaser, SQLRight, and WindRanger when testing DBMSs using testing oracles. For NoREC and TLP oracles, they are implemented in tested four DBMSs.
For ROWID, LIKELY, and INDEX oracles, they are only implemented in SQLite for now, and not in other DBMSs yet. 

According to Fig.~\ref{fig:3}, Fig.~\ref{fig:4} and Fig.~\ref{fig:5}, we can see that SQLaser has a lower code coverage compared to SQLRight. This is because the additional coverage of SQLRight primarily comes from its coverage-guided exploration of code paths, while SQLaser employs a targeted fuzzing strategy. However, despite having higher coverage within the same time frame, SQLRight does not discover more logic bugs. This demonstrates that blind path exploration is not an ideal strategy, as many of the explored paths do not trigger logic bugs.
We can also observe that in terms of code coverage, SQLaser tends to be slightly lower than WindRanger. This is because
SQLaser's target sites are function call chains that characterize patterns of logic bugs. WindRanger's target sites are set to the individual functions where patches for existing logic bugs are located. Consequently, targeting function call chains slightly constrains the exploration capability of the fuzzer compared to individual functions, resulting in a marginally lower code coverage of SQLaser than WindRanger.

\subsubsection{Seed Distance Distribution} \label{Seed Distance Distribution}
\begin{figure*}[ht]
\centering
\includegraphics[width=0.8\linewidth]{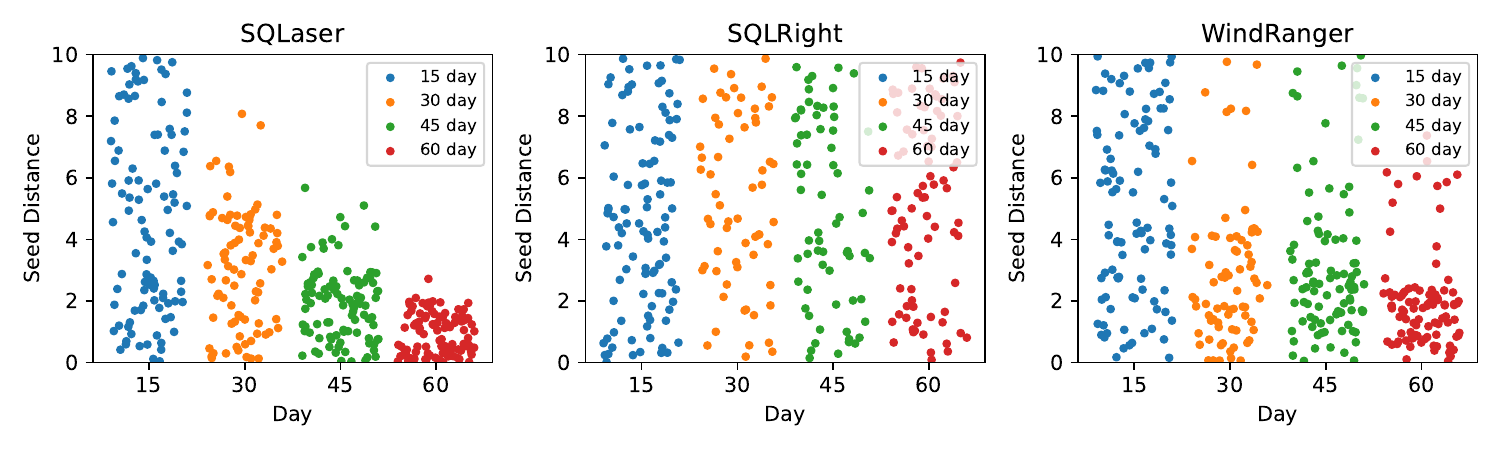}
\caption{Seed distance distribution of SQLaser, SQLRight, and WindRanger testing SQLite. }
\label{fig:6}
\end{figure*}

To assess the evolution of seed distances from the target call chain over time, we conduct tests on SQLite, randomly selecting 100 seeds at 15, 30, 45, and 60 days and showcasing their distances from the target call chains. As illustrated in Fig.~\ref{fig:6}, for SQLaser, at day 15, some seeds are still relatively far from the target call chain, and the distribution of seed distances is quite dispersed. However, as the testing period increases, seed distances gradually decrease, with most seeds converging to values between 0 and 2 by day 60. In contrast, SQLRight, lacking a feedback strategy based on seed distances from the target call chain, exhibits consistently dispersed seed distances over the 60-day period. WindRanger, focusing on a singular function rather than the entire call chain, experiences some convergence in seed distances by day 60, but a portion of seeds still exhibit larger distances. This highlights the efficacy of our distance calculation algorithm, showcasing that seeds generated by SQLaser become increasingly closer to the target call chains over time.

\subsection{Trimmed Call Chains vs.\ Complete Call Chains} \label{Trimmed Complete}
In this paper, we set the target of SQLaser to be a trimmed call chain rather than a complete call chain, aiming to enhance the fuzzer's capability to discover more new bugs. In this section, we compare the results between trimmed call chain and complete call chain, including the false positives and false negatives introduced by these two approaches. To achieve this goal, we additionally configure SQLaser's target as a complete call chain and conduct a 30-day test on SQLite.
\begin{table}[]
\caption{During the 30-day testing of SQLite with SQLaser targeting trimmed call chains (SQLaser\_trim) and complete call chains (SQLaser\_complete), the number of false positives and finding bugs.}
\label{tab:completeCallChain}
\footnotesize
\centering
\begin{tabular}{ccc}
\hline
                 & \bf{\emph{SQLaser\_trim }}& \bf{\emph{SQLaser\_complete}} \\ \hline
False Positives   & 25            & 18      \\
Finding Bugs & 8             & 3                 \\
\hline

\end{tabular}
\end{table}

As shown in Table~\ref{tab:completeCallChain}, when the target of the fuzzer is trimmed call chains, SQLaser produces 25 false positives in SQLite (Manual verification of false positives is not a time-consuming task. Our master student spent only half an hour to validate 25 false positives). On the other hand, when the target is complete call chains, SQLaser reports 18 false positives in SQLite. Additionally, targeting trimmed call chains has led to finding 8 bugs, whereas targeting complete call chains has resulted in SQLaser finding 3 bugs. While we may not have information about the total number of logic bugs, based on these results, it can be anticipated that targeting trimmed call chains would exhibit fewer false negatives. This is suggested by the higher number of bug reproductions when focusing on trimmed call chains.

Therefore, within the 30-day timeframe, 
when targeting trimmed call chains, SQLaser generates more false positives because it explores a broader range of code paths. It may encounter situations where the reported errors are not actual bugs.
On the other hand, targeting complete call chains reduces the false positives as the fuzzer has a better understanding of the target's entire context.
However, targeting on complete call chains may result in more false negatives, because it may avoid certain paths that could lead to bugs. Since it has a more specific target of the intended behavior, it will constrain the exploration capability of fuzzing.
Targeting trimmed call chains (which represent a subset of the complete execution) may lead to the discovery of more bugs, since the fuzzer explores a wider range of possibilities, even if some of them are false positives.

\section{Discussion}
In this section, we discuss two issues: the completeness of the testing oracles employed in this paper, and the completeness of the SQL clause collections we consider.

\subsection{Completeness of Testing Oracles} \label{Completeness of Testing Oracles}
In this paper, we employ five oracles to test logic bugs. Specifically, we validate the existence of logic bugs by converting SQL query statements into functionally equivalent yet different queries and comparing the output results of these two queries. Currently, SQLaser only supports differential testing-based oracles and does not accommodate other types of oracles, such as PQS oracle~\cite{PQS}, which detects logic bugs through non-differential testing. However, in the process of statistically analyzing existing logic bugs, we include the bugs detected by PQS oracles and instrument their call chains. As SQLaser does not currently support PQS oracles, this may result in its inability to identify logic bugs detectable solely by PQS oracles. Additionally, to the best of our knowledge, existing research has not extensively investigated the completeness of these five oracles (NoREC, TLP, INDEX, ROWID and LIKELY) in addressing logic bugs in real-world scenarios. We will undertake research on this matter in future work.

\subsection{Completeness of SQL Clause Collections}
We systematically analyze logic bugs extracted from state-of-the-art studies, identifying common error-prone sequences of functions. We have made every effort to comprehensively gather existing logic bugs, and these resultant collections effectively represent prevalent logic bugs in the field. To our best knowledge, there is currently no theory or standard proposed in existing research to classify SQL clauses. We are the first to strive to investigate the types of SQL clause of existing logic bugs, and leverage the discovered insights for detecting bugs in DBMS. However, we acknowledge that ensuring this classification covers all SQL clause might be a challenge, and addressing this limitation is a direction for our future work.

\section{Related Work}
In this section, we discuss techniques for detecting logic bugs in DBMSs. SQLancer~\cite{SQLancer} is a rule-based tool that successfully identifies numerous logic bugs. It employs three distinct methods for detecting logic bugs: PQS~\cite{PQS}, NoREC~\cite{NoREC}, and TLP~\cite{TLP}. PQS focuses on generating queries that fetch a randomly selected row called the pivot row. If the DBMS fails to retrieve the pivot row based on this random expression, it indicates the presence of a logic bug in the DBMS. NoREC, on the other hand, translates an optimized query into an unoptimized version with identical semantics. By comparing the results obtained from both queries, inconsistencies can indicate the presence of a logic bug. Lastly, TLP partitions a query into several sub-queries, which are composed to have the same semantics as the original query. If the composed query does not yield the same result as the original query, it suggests the possibility of a logic bug.

While SQLancer relies on a rule-based generator to detect logic bugs, limiting its exploration of input samples. Liang~\etal~\cite{SQLRight} propose SQLRight, a methodology for logic bug exploration based on coverage-based fuzzing. SQLRight successfully detects logic bugs in SQLite and MySQL. In their approach, all sample queries are added to a queue, and the fuzzer selects a query to mutate. If the mutated query triggers a new execution path, it is added to the queue. These mutated queries are then tested to identify any unexpected behaviors.
However, SQLRight's coverage-based fuzzing approach lacks high effectiveness in reaching vulnerable target sites due to the equal evaluation of all seeds. In this paper, we propose SQLaser,
a SQL-clause-guided fuzzing approach, which focuses primarily on 
seeds which follow certain bug patterns—involving a combination of the effects of multiple SQL clauses. Through our analysis, behind these clause combinations are a sequence of functions. We thus use these function chains (\ie, sequence of functions) as target sites. Our evaluation shows that SQLaser significantly reduces the bug discovery time.

\section{Conclusion}
This paper contributes by analyzing existing logic bugs and identifying 35 SQL-level bug patterns associated with SQL clause combinations that result in logic bugs. We treat function chains related to these SQL clause combinations as the target sites of directed fuzzing. To validate whether these target function chains can cause more logic bugs, we propose SQLaser which employs directed fuzzing to efficiently reach the target sites. Notably, we propose a new algorithm to compute the distance between seed call chains and the target call chains, in order to prioritize the seeds with shorter distances.  
We evaluate SQLaser to assess its effectiveness and efficiency in detecting logic bugs in DBMSs. The approach is validated on SQLite, MySQL, PostgreSQL and TiDB, and successfully uncover 4 new unique logic bugs (verified by vendors). Furthermore, SQLaser outperforms SQLRight and WindRanger in terms of efficiency, with a reduction of at least 60\% in bug discovery time.





\nocite{*}
\bibliographystyle{ios1}           
\bibliography{bibliography}        

\begin{thebibliography}{97}
\ifx \bisbn   \undefined \def \bisbn  #1{ISBN #1}\fi
\ifx \binits  \undefined \def \binits#1{#1} \fi
\ifx \bauthor  \undefined \def \bauthor#1{#1} \fi
\ifx \bjtitle  \undefined \def \bjtitle#1{\textit{#1}}\fi
\ifx \batitle  \undefined \def \batitle#1{#1} \fi
\ifx \bctitle  \undefined \def \bctitle#1{#1} \fi
\ifx \bvolume  \undefined \def \bvolume#1{\textbf{#1}}\fi
\ifx \byear  \undefined \def \byear#1{#1} \fi
\ifx \bissue  \undefined \def \bissue#1{#1} \fi
\ifx \bfpage  \undefined \def \bfpage#1{#1} \fi
\ifx \blpage  \undefined \def \blpage #1{#1} \fi
\ifx \burl  \undefined \def \burl#1{#1} \fi
\ifx \doiurl  \undefined \def \doiurl#1{#1} \fi
\ifx \betal  \undefined \def \betal{et al.} \fi
\ifx \binstitute  \undefined \def \binstitute#1{#1} \fi
\ifx \beditor  \undefined \def \beditor#1{#1} \fi
\ifx \bpublisher  \undefined \def \bpublisher#1{#1} \fi
\ifx \bbtitle  \undefined \def \bbtitle#1{\textit{#1}} \fi
\ifx \bedition  \undefined \def \bedition#1{#1} \fi
\ifx \bseriesno  \undefined \def \bseriesno#1{#1} \fi
\ifx \blocation  \undefined \def \blocation#1{#1} \fi
\ifx \bsertitle  \undefined \def \bsertitle#1{#1} \fi
\ifx \bsnm \undefined \def \bsnm#1{#1} \fi
\ifx \bsuffix \undefined \def \bsuffix#1{#1} \fi
\ifx \bparticle \undefined \def \bparticle#1{#1} \fi
\ifx \barticle \undefined \def \barticle#1{#1} \fi
\ifx \botherref \undefined \def \botherref #1{#1} \fi
\ifx \url \undefined \def \url#1{#1} \fi
\ifx \bchapter \undefined \def \bchapter#1{#1} \fi
\ifx \bbook \undefined \def \bbook#1{#1} \fi
\ifx \bcomment \undefined \def \bcomment#1{#1} \fi
\ifx \oauthor \undefined \def \oauthor#1{#1} \fi
\ifx \citeauthoryear \undefined \def \citeauthoryear#1{#1} \fi
\ifx \texttildelow  \undefined \def \texttildelow{\symbol{126}} \fi
\def \endbibitem {}
\ifx \bconflocation  \undefined \def \bconflocation#1{#1} \fi

\bibitem{SQLquery}
\begin{bchapter}
\bauthor{\binits{K.S.}~\bsnm{Abdul}} and
\bauthor{\binits{S.}~\bsnm{Khurshid}},
\bctitle{Automated SQL query generation for systematic testing of database engines},
in: \bbtitle{Proceedings of the 25th IEEE/ACM International Conference on Automated Software Engineering},
\byear{2010},
pp.~\bfpage{329}--\blpage{332}.
\end{bchapter}
\endbibitem

\bibitem{REDQUEEN}
\begin{bchapter}
\bauthor{\binits{C.}~\bsnm{Aschermann}},
\bauthor{\binits{S.}~\bsnm{Schumilo}},
\bauthor{\binits{T.}~\bsnm{Blazytko}},
\bauthor{\binits{R.}~\bsnm{Gawlik}} and
\bauthor{\binits{T.}~\bsnm{Holz}},
\bctitle{REDQUEEN: Fuzzing with Input-to-State Correspondence.},
in: \bbtitle{Proceedings of the 26th Annual Network and Distributed System Security Symposium (NDSS)},
Vol.~\bseriesno{19},
\byear{2019},
pp.~\bfpage{1}--\blpage{15}.
\end{bchapter}
\endbibitem

\bibitem{GRIMOIRE}
\begin{bchapter}
\bauthor{\binits{T.}~\bsnm{Blazytko}},
\bauthor{\binits{M.}~\bsnm{Bishop}},
\bauthor{\binits{C.}~\bsnm{Aschermann}},
\bauthor{\binits{J.}~\bsnm{Cappos}},
\bauthor{\binits{M.}~\bsnm{Schl{\"o}gel}},
\bauthor{\binits{N.}~\bsnm{Korshun}},
\bauthor{\binits{A.}~\bsnm{Abbasi}},
\bauthor{\binits{M.}~\bsnm{Schweighauser}},
\bauthor{\binits{S.}~\bsnm{Schinzel}},
\bauthor{\binits{S.}~\bsnm{Schumilo}} \betal,
\bctitle{$\{$GRIMOIRE$\}$: Synthesizing structure while fuzzing},
in: \bbtitle{Proceedings of the 28th USENIX Security Symposium (USENIX Security 19)},
\byear{2019},
pp.~\bfpage{1985}--\blpage{2002}.
\end{bchapter}
\endbibitem

\bibitem{AFLGo}
\begin{bchapter}
\bauthor{\binits{M.}~\bsnm{B{\"o}hme}},
\bauthor{\binits{V.-T.}~\bsnm{Pham}},
\bauthor{\binits{M.-D.}~\bsnm{Nguyen}} and
\bauthor{\binits{A.}~\bsnm{Roychoudhury}},
\bctitle{Directed greybox fuzzing},
in: \bbtitle{Proceedings of the 2017 ACM SIGSAC conference on computer and communications security},
\byear{2017},
pp.~\bfpage{2329}--\blpage{2344}.
\end{bchapter}
\endbibitem

\bibitem{Hawkeye}
\begin{bchapter}
\bauthor{\binits{H.}~\bsnm{Chen}},
\bauthor{\binits{Y.}~\bsnm{Xue}},
\bauthor{\binits{Y.}~\bsnm{Li}},
\bauthor{\binits{B.}~\bsnm{Chen}},
\bauthor{\binits{X.}~\bsnm{Xie}},
\bauthor{\binits{X.}~\bsnm{Wu}} and
\bauthor{\binits{Y.}~\bsnm{Liu}},
\bctitle{Hawkeye: Towards a desired directed grey-box fuzzer},
in: \bbtitle{Proceedings of the 2018 ACM SIGSAC conference on computer and communications security},
\byear{2018},
pp.~\bfpage{2095}--\blpage{2108}.
\end{bchapter}
\endbibitem

\bibitem{Angora}
\begin{bchapter}
\bauthor{\binits{P.}~\bsnm{Chen}} and
\bauthor{\binits{H.}~\bsnm{Chen}},
\bctitle{Angora: Efficient fuzzing by principled search},
in: \bbtitle{Proceedings of the 2018 IEEE Symposium on Security and Privacy (SP)},
\binstitute{IEEE},
\byear{2018},
pp.~\bfpage{711}--\blpage{725}.
\end{bchapter}
\endbibitem

\bibitem{OneEngine}
\begin{bchapter}
\bauthor{\binits{Y.}~\bsnm{Chen}},
\bauthor{\binits{R.}~\bsnm{Zhong}},
\bauthor{\binits{H.}~\bsnm{Hu}},
\bauthor{\binits{H.}~\bsnm{Zhang}},
\bauthor{\binits{Y.}~\bsnm{Yang}},
\bauthor{\binits{D.}~\bsnm{Wu}} and
\bauthor{\binits{W.}~\bsnm{Lee}},
\bctitle{One engine to fuzz’em all: Generic language processor testing with semantic validation},
in: \bbtitle{Proceedings of the 2021 IEEE Symposium on Security and Privacy (SP)},
\binstitute{IEEE},
\byear{2021},
pp.~\bfpage{642}--\blpage{658}.
\end{bchapter}
\endbibitem

\bibitem{NTFUZZ}
\begin{bchapter}
\bauthor{\binits{J.}~\bsnm{Choi}},
\bauthor{\binits{K.}~\bsnm{Kim}},
\bauthor{\binits{D.}~\bsnm{Lee}} and
\bauthor{\binits{S.K.}~\bsnm{Cha}},
\bctitle{NTFuzz: Enabling type-aware kernel fuzzing on windows with static binary analysis},
in: \bbtitle{Proceedings of the 2021 IEEE Symposium on Security and Privacy (SP)},
\binstitute{IEEE},
\byear{2021},
pp.~\bfpage{677}--\blpage{693}.
\end{bchapter}
\endbibitem

\bibitem{fuzzing}
\begin{bchapter}
\bauthor{\binits{Z.Y.}~\bsnm{Ding}} and
\bauthor{\binits{C.L.}~\bsnm{Goues}},
\bctitle{An empirical study of oss-fuzz bugs},
in: \bbtitle{Proceedings of the 2021 IEEE/ACM 18th International Conference on Mining Software Repositories (MSR)},
\binstitute{IEEE},
\byear{2021},
pp.~\bfpage{131}--\blpage{142}.
\end{bchapter}
\endbibitem

\bibitem{Favocado}
\begin{bchapter}
\bauthor{\binits{S.T.}~\bsnm{Dinh}},
\bauthor{\binits{H.}~\bsnm{Cho}},
\bauthor{\binits{K.}~\bsnm{Martin}},
\bauthor{\binits{A.}~\bsnm{Oest}},
\bauthor{\binits{K.}~\bsnm{Zeng}},
\bauthor{\binits{A.}~\bsnm{Kapravelos}},
\bauthor{\binits{G.-J.}~\bsnm{Ahn}},
\bauthor{\binits{T.}~\bsnm{Bao}},
\bauthor{\binits{R.}~\bsnm{Wang}},
\bauthor{\binits{A.}~\bsnm{Doup{\'e}}} \betal,
\bctitle{Favocado: Fuzzing the Binding Code of JavaScript Engines Using Semantically Correct Test Cases.},
in: \bbtitle{Proceedings of the 28th Annual Network and Distributed System Security Symposium (NDSS)},
\byear{2021}.
\end{bchapter}
\endbibitem

\bibitem{Leopard}
\begin{bchapter}
\bauthor{\binits{X.}~\bsnm{Du}},
\bauthor{\binits{B.}~\bsnm{Chen}},
\bauthor{\binits{Y.}~\bsnm{Li}},
\bauthor{\binits{J.}~\bsnm{Guo}},
\bauthor{\binits{Y.}~\bsnm{Zhou}},
\bauthor{\binits{Y.}~\bsnm{Liu}} and
\bauthor{\binits{Y.}~\bsnm{Jiang}},
\bctitle{Leopard: Identifying vulnerable code for vulnerability assessment through program metrics},
in: \bbtitle{Proceedings of the 2019 IEEE/ACM 41st International Conference on Software Engineering (ICSE)},
\binstitute{IEEE},
\byear{2019},
pp.~\bfpage{60}--\blpage{71}.
\end{bchapter}
\endbibitem

\bibitem{WindRanger}
\begin{bchapter}
\bauthor{\binits{Z.}~\bsnm{Du}},
\bauthor{\binits{Y.}~\bsnm{Li}},
\bauthor{\binits{Y.}~\bsnm{Liu}} and
\bauthor{\binits{B.}~\bsnm{Mao}},
\bctitle{WindRanger: a directed greybox fuzzer driven by deviation basic blocks},
in: \bbtitle{Proceedings of the 44th International Conference on Software Engineering},
\byear{2022},
pp.~\bfpage{2440}--\blpage{2451}.
\end{bchapter}
\endbibitem

\bibitem{DBMS}
\begin{barticle}
\bauthor{\binits{J.K.}~\bsnm{Fichte}},
\bauthor{\binits{M.}~\bsnm{Hecher}},
\bauthor{\binits{P.}~\bsnm{Thier}} and
\bauthor{\binits{S.}~\bsnm{Woltran}},
\batitle{Exploiting database management systems and treewidth for counting},
\bjtitle{Theory and Practice of Logic Programming}
\bvolume{22}(\bissue{1})
(\byear{2022}),
\bfpage{128}--\blpage{157}.
\end{barticle}
\endbibitem

\bibitem{TiDB}
\begin{botherref}
\oauthor{\binits{P.}~\bsnm{Foundation}},
TiDB,
2023.
\url{https://www.pingcap.com/}.
\end{botherref}
\endbibitem

\bibitem{GREYONE}
\begin{bchapter}
\bauthor{\binits{S.}~\bsnm{Gan}},
\bauthor{\binits{C.}~\bsnm{Zhang}},
\bauthor{\binits{P.}~\bsnm{Chen}},
\bauthor{\binits{B.}~\bsnm{Zhao}},
\bauthor{\binits{X.}~\bsnm{Qin}},
\bauthor{\binits{D.}~\bsnm{Wu}} and
\bauthor{\binits{Z.}~\bsnm{Chen}},
\bctitle{GREYONE: Data flow sensitive fuzzing},
in: \bbtitle{Proceedings of the 29th USENIX security symposium (USENIX Security 20)},
\byear{2020},
pp.~\bfpage{2577}--\blpage{2594}.
\end{bchapter}
\endbibitem

\bibitem{ghit2020sparkfuzz}
\begin{bchapter}
\bauthor{\binits{B.}~\bsnm{Ghit}},
\bauthor{\binits{N.}~\bsnm{Poggi}},
\bauthor{\binits{J.}~\bsnm{Rosen}},
\bauthor{\binits{R.}~\bsnm{Xin}} and
\bauthor{\binits{P.}~\bsnm{Boncz}},
\bctitle{SparkFuzz: Searching correctness regressions in modern query engines},
in: \bbtitle{Proceedings of the workshop on Testing Database Systems},
\byear{2020},
pp.~\bfpage{1}--\blpage{6}.
\end{bchapter}
\endbibitem

\bibitem{ClusterFuzz}
\begin{botherref}
\oauthor{\bsnm{Google}},
ClusterFuzz,
2023.
\url{https://google.github.io/clusterfuzz}.
\end{botherref}
\endbibitem

\bibitem{Honggfuzz}
\begin{botherref}
\oauthor{\bsnm{Google}},
Honggfuzz,
2023.
\url{https://google.github.io/honggfuzz/}.
\end{botherref}
\endbibitem

\bibitem{PostgreSQL}
\begin{botherref}
\oauthor{\binits{P.D.}~\bsnm{Group}},
PostgreSQL,
2023.
\url{https://www.postgresql.org/}.
\end{botherref}
\endbibitem

\bibitem{CodeAlchemist}
\begin{bchapter}
\bauthor{\binits{H.}~\bsnm{Han}},
\bauthor{\binits{D.}~\bsnm{Oh}} and
\bauthor{\binits{S.K.}~\bsnm{Cha}},
\bctitle{CodeAlchemist: Semantics-Aware Code Generation to Find Vulnerabilities in JavaScript Engines.},
in: \bbtitle{Proceedings of the 26th Network and Distributed System Security Symposium (NDSS)},
\byear{2019}.
\end{bchapter}
\endbibitem

\bibitem{PIN}
\begin{botherref}
\oauthor{\bsnm{Intel}},
PIN,
2023.
\url{https://www.intel.cn/content/www/cn/zh/developer/articles/tool/pin-a-dynamic-binary-instrumentation-tool.html}.
\end{botherref}
\endbibitem

\bibitem{ContractFuzzer}
\begin{bchapter}
\bauthor{\binits{B.}~\bsnm{Jiang}},
\bauthor{\binits{Y.}~\bsnm{Liu}} and
\bauthor{\binits{W.K.}~\bsnm{Chan}},
\bctitle{Contractfuzzer: Fuzzing smart contracts for vulnerability detection},
in: \bbtitle{Proceedings of the 33rd ACM/IEEE International Conference on Automated Software Engineering},
\byear{2018},
pp.~\bfpage{259}--\blpage{269}.
\end{bchapter}
\endbibitem

\bibitem{APOLLO}
\begin{barticle}
\bauthor{\binits{J.}~\bsnm{Jung}},
\bauthor{\binits{H.}~\bsnm{Hu}},
\bauthor{\binits{J.}~\bsnm{Arulraj}},
\bauthor{\binits{T.}~\bsnm{Kim}} and
\bauthor{\binits{W.}~\bsnm{Kang}},
\batitle{Apollo: Automatic detection and diagnosis of performance regressions in database systems},
\bjtitle{Proceedings of the 46th International Conference on Very Large Data Bases (VLDB)}
\bvolume{13}(\bissue{1})
(\byear{2019}),
\bfpage{57}--\blpage{70}.
\end{barticle}
\endbibitem

\bibitem{HFL}
\begin{bchapter}
\bauthor{\binits{K.}~\bsnm{Kim}},
\bauthor{\binits{D.R.}~\bsnm{Jeong}},
\bauthor{\binits{C.H.}~\bsnm{Kim}},
\bauthor{\binits{Y.}~\bsnm{Jang}},
\bauthor{\binits{I.}~\bsnm{Shin}} and
\bauthor{\binits{B.}~\bsnm{Lee}},
\bctitle{HFL: Hybrid Fuzzing on the Linux Kernel.},
in: \bbtitle{Proceedings of the 27th Annual Network and Distributed System Security Symposium (NDSS)},
\byear{2020}.
\end{bchapter}
\endbibitem

\bibitem{Lee}
\begin{bchapter}
\bauthor{\binits{G.}~\bsnm{Lee}},
\bauthor{\binits{W.}~\bsnm{Shim}} and
\bauthor{\binits{B.}~\bsnm{Lee}},
\bctitle{Constraint-guided directed greybox fuzzing},
in: \bbtitle{Proceedings of the 30th USENIX Security Symposium (USENIX Security 21)},
\byear{2021},
pp.~\bfpage{3559}--\blpage{3576}.
\end{bchapter}
\endbibitem

\bibitem{Fairfuzz}
\begin{bchapter}
\bauthor{\binits{C.}~\bsnm{Lemieux}} and
\bauthor{\binits{K.}~\bsnm{Sen}},
\bctitle{Fairfuzz: A targeted mutation strategy for increasing greybox fuzz testing coverage},
in: \bbtitle{Proceedings of the 33rd ACM/IEEE international conference on automated software engineering},
\byear{2018},
pp.~\bfpage{475}--\blpage{485}.
\end{bchapter}
\endbibitem

\bibitem{Steelix}
\begin{bchapter}
\bauthor{\binits{Y.}~\bsnm{Li}},
\bauthor{\binits{B.}~\bsnm{Chen}},
\bauthor{\binits{M.}~\bsnm{Chandramohan}},
\bauthor{\binits{S.-W.}~\bsnm{Lin}},
\bauthor{\binits{Y.}~\bsnm{Liu}} and
\bauthor{\binits{A.}~\bsnm{Tiu}},
\bctitle{Steelix: program-state based binary fuzzing},
in: \bbtitle{Proceedings of the 2017 11th joint meeting on foundations of software engineering},
\byear{2017},
pp.~\bfpage{627}--\blpage{637}.
\end{bchapter}
\endbibitem

\bibitem{SQLRight}
\begin{bchapter}
\bauthor{\binits{Y.}~\bsnm{Liang}},
\bauthor{\binits{S.}~\bsnm{Liu}} and
\bauthor{\binits{H.}~\bsnm{Hu}},
\bctitle{Detecting Logical Bugs of $\{$DBMS$\}$ with Coverage-based Guidance},
in: \bbtitle{Proceedings of the 31st USENIX Security Symposium (USENIX Security 22)},
\byear{2022},
pp.~\bfpage{4309}--\blpage{4326}.
\end{bchapter}
\endbibitem

\bibitem{LiuX}
\begin{botherref}
\oauthor{\binits{X.}~\bsnm{Liu}},
\oauthor{\binits{Q.}~\bsnm{Zhou}},
\oauthor{\binits{J.}~\bsnm{Arulraj}} and
\oauthor{\binits{A.}~\bsnm{Orso}},
Automated performance bug detection in database systems,
\textit{arXiv preprint arXiv:2105.10016}
(2021).
\end{botherref}
\endbibitem

\bibitem{LibFuzzer}
\begin{botherref}
\oauthor{\bsnm{LLVM}},
LibFuzzer: A Library For Coverage-guided Fuzz Testing,
2023.
\url{http://llvm.org/docs/LibFuzzer.html}.
\end{botherref}
\endbibitem

\bibitem{lo2010framework}
\begin{barticle}
\bauthor{\binits{E.}~\bsnm{Lo}},
\bauthor{\binits{C.}~\bsnm{Binnig}},
\bauthor{\binits{D.}~\bsnm{Kossmann}},
\bauthor{\binits{M.}~\bsnm{Tamer~{\"O}zsu}} and
\bauthor{\binits{W.-K.}~\bsnm{Hon}},
\batitle{A framework for testing DBMS features},
\bjtitle{The VLDB Journal}
\bvolume{19}
(\byear{2010}),
\bfpage{203}--\blpage{230}.
\end{barticle}
\endbibitem

\bibitem{mckeeman1998differential}
\begin{barticle}
\bauthor{\binits{W.M.}~\bsnm{McKeeman}},
\batitle{Differential testing for software},
\bjtitle{Digital Technical Journal}
\bvolume{10}(\bissue{1})
(\byear{1998}),
\bfpage{100}--\blpage{107}.
\end{barticle}
\endbibitem

\bibitem{MySQL}
\begin{botherref}
\oauthor{\bsnm{MySQL}},
MySQL Customers,
2023.
\url{https://www.mysql.com/}.
\end{botherref}
\endbibitem

\bibitem{sfuzz}
\begin{bchapter}
\bauthor{\binits{T.D.}~\bsnm{Nguyen}},
\bauthor{\binits{L.H.}~\bsnm{Pham}},
\bauthor{\binits{J.}~\bsnm{Sun}},
\bauthor{\binits{Y.}~\bsnm{Lin}} and
\bauthor{\binits{Q.T.}~\bsnm{Minh}},
\bctitle{sfuzz: An efficient adaptive fuzzer for solidity smart contracts},
in: \bbtitle{Proceedings of the ACM/IEEE 42nd International Conference on Software Engineering},
\byear{2020},
pp.~\bfpage{778}--\blpage{788}.
\end{bchapter}
\endbibitem

\bibitem{Parmesan}
\begin{bchapter}
\bauthor{\binits{S.}~\bsnm{{\"O}sterlund}},
\bauthor{\binits{K.}~\bsnm{Razavi}},
\bauthor{\binits{H.}~\bsnm{Bos}} and
\bauthor{\binits{C.}~\bsnm{Giuffrida}},
\bctitle{$\{$ParmeSan$\}$: Sanitizer-guided greybox fuzzing},
in: \bbtitle{Proceedings of the 29th USENIX Security Symposium (USENIX Security 20)},
\byear{2020},
pp.~\bfpage{2289}--\blpage{2306}.
\end{bchapter}
\endbibitem

\bibitem{MoonShine}
\begin{bchapter}
\bauthor{\binits{S.}~\bsnm{Pailoor}},
\bauthor{\binits{A.}~\bsnm{Aday}} and
\bauthor{\binits{S.}~\bsnm{Jana}},
\bctitle{$\{$MoonShine$\}$: Optimizing $\{$OS$\}$ fuzzer seed selection with trace distillation},
in: \bbtitle{Proceedings of the 27th USENIX Security Symposium (USENIX Security 18)},
\byear{2018},
pp.~\bfpage{729}--\blpage{743}.
\end{bchapter}
\endbibitem

\bibitem{FuzzingJavaScrip}
\begin{bchapter}
\bauthor{\binits{S.}~\bsnm{Park}},
\bauthor{\binits{W.}~\bsnm{Xu}},
\bauthor{\binits{I.}~\bsnm{Yun}},
\bauthor{\binits{D.}~\bsnm{Jang}} and
\bauthor{\binits{T.}~\bsnm{Kim}},
\bctitle{Fuzzing javascript engines with aspect-preserving mutation},
in: \bbtitle{Proceedings of the 2020 IEEE Symposium on Security and Privacy (SP)},
\binstitute{IEEE},
\byear{2020},
pp.~\bfpage{1629}--\blpage{1642}.
\end{bchapter}
\endbibitem

\bibitem{NoREC}
\begin{bchapter}
\bauthor{\binits{M.}~\bsnm{Rigger}} and
\bauthor{\binits{Z.}~\bsnm{Su}},
\bctitle{Detecting optimization bugs in database engines via non-optimizing reference engine construction},
in: \bbtitle{Proceedings of the 28th ACM Joint Meeting on European Software Engineering Conference and Symposium on the Foundations of Software Engineering},
\byear{2020},
pp.~\bfpage{1140}--\blpage{1152}.
\end{bchapter}
\endbibitem

\bibitem{TLP}
\begin{bchapter}
\bauthor{\binits{M.}~\bsnm{Rigger}} and
\bauthor{\binits{Z.}~\bsnm{Su}},
\bctitle{Finding bugs in database systems via query partitioning},
in: \bbtitle{Proceedings of the ACM on Programming Languages},
Vol.~\bseriesno{4},
\bpublisher{ACM New York, NY, USA},
\byear{2020},
pp.~\bfpage{1}--\blpage{30}.
\end{bchapter}
\endbibitem

\bibitem{PQS}
\begin{bchapter}
\bauthor{\binits{M.}~\bsnm{Rigger}} and
\bauthor{\binits{Z.}~\bsnm{Su}},
\bctitle{Testing database engines via pivoted query synthesis},
in: \bbtitle{Proceedings of 14th USENIX Symposium on Operating Systems Design and Implementation (OSDI 20)},
\byear{2020},
pp.~\bfpage{667}--\blpage{682}.
\end{bchapter}
\endbibitem

\bibitem{Serebryany}
\begin{botherref}
\oauthor{\binits{K.}~\bsnm{Serebryany}},
Sanitize, fuzz, and harden your C++ code,
\textit{San Francisco, CA}
(2016).
\end{botherref}
\endbibitem

\bibitem{slutz1998massive}
\begin{bchapter}
\bauthor{\binits{D.R.}~\bsnm{Slutz}},
\bctitle{Massive stochastic testing of SQL},
in: \bbtitle{VLDB},
Vol.~\bseriesno{98},
\binstitute{Citeseer},
\byear{1998},
pp.~\bfpage{618}--\blpage{622}.
\end{bchapter}
\endbibitem

\bibitem{SQLite}
\begin{botherref}
\oauthor{\bsnm{SQLite}},
SQLite,
2023.
\url{https://sqlite.org/index.html}.
\end{botherref}
\endbibitem

\bibitem{bugfix}
\begin{botherref}
\oauthor{\bsnm{SQLite}},
SQLite bug fix,
2023.
\url{https://www.sqlite.org/src/info/5351e920f489562f}.
\end{botherref}
\endbibitem

\bibitem{Driller}
\begin{bchapter}
\bauthor{\binits{N.}~\bsnm{Stephens}},
\bauthor{\binits{J.}~\bsnm{Grosen}},
\bauthor{\binits{C.}~\bsnm{Salls}},
\bauthor{\binits{A.}~\bsnm{Dutcher}},
\bauthor{\binits{R.}~\bsnm{Wang}},
\bauthor{\binits{J.}~\bsnm{Corbetta}},
\bauthor{\binits{Y.}~\bsnm{Shoshitaishvili}},
\bauthor{\binits{C.}~\bsnm{Kruegel}} and
\bauthor{\binits{G.}~\bsnm{Vigna}},
\bctitle{Driller: Augmenting fuzzing through selective symbolic execution.},
in: \bbtitle{NDSS},
Vol.~\bseriesno{16},
\byear{2016},
pp.~\bfpage{1}--\blpage{16}.
\end{bchapter}
\endbibitem

\bibitem{WangM}
\begin{bchapter}
\bauthor{\binits{M.}~\bsnm{Wang}},
\bauthor{\binits{Z.}~\bsnm{Wu}},
\bauthor{\binits{X.}~\bsnm{Xu}},
\bauthor{\binits{J.}~\bsnm{Liang}},
\bauthor{\binits{C.}~\bsnm{Zhou}},
\bauthor{\binits{H.}~\bsnm{Zhang}} and
\bauthor{\binits{Y.}~\bsnm{Jiang}},
\bctitle{Industry practice of coverage-guided enterprise-level DBMS fuzzing},
in: \bbtitle{Proceedings of the 2021 IEEE/ACM 43rd International Conference on Software Engineering: Software Engineering in Practice (ICSE-SEIP)},
\binstitute{IEEE},
\byear{2021},
pp.~\bfpage{328}--\blpage{337}.
\end{bchapter}
\endbibitem

\bibitem{ASCII}
\begin{botherref}
\oauthor{\bsnm{wikipedia}},
ASCII,
2023.
\url{https://zh.wikipedia.org/wiki/ASCII}.
\end{botherref}
\endbibitem

\bibitem{coredump}
\begin{botherref}
\oauthor{\bsnm{wikipedia}},
coredump,
2023.
\url{https://en.wikipedia.org/wiki/Core_dump}.
\end{botherref}
\endbibitem

\bibitem{Harvey}
\begin{bchapter}
\bauthor{\binits{V.}~\bsnm{W{\"u}stholz}} and
\bauthor{\binits{M.}~\bsnm{Christakis}},
\bctitle{Harvey: A greybox fuzzer for smart contracts},
in: \bbtitle{Proceedings of the 28th ACM Joint Meeting on European Software Engineering Conference and Symposium on the Foundations of Software Engineering},
\byear{2020},
pp.~\bfpage{1398}--\blpage{1409}.
\end{bchapter}
\endbibitem

\bibitem{Krace}
\begin{bchapter}
\bauthor{\binits{M.}~\bsnm{Xu}},
\bauthor{\binits{S.}~\bsnm{Kashyap}},
\bauthor{\binits{H.}~\bsnm{Zhao}} and
\bauthor{\binits{T.}~\bsnm{Kim}},
\bctitle{Krace: Data race fuzzing for kernel file systems},
in: \bbtitle{Proceedings of the 2020 IEEE Symposium on Security and Privacy (SP)},
\binstitute{IEEE},
\byear{2020},
pp.~\bfpage{1643}--\blpage{1660}.
\end{bchapter}
\endbibitem

\bibitem{Cooper}
\begin{bchapter}
\bauthor{\binits{P.}~\bsnm{Xu}},
\bauthor{\binits{Y.}~\bsnm{Wang}},
\bauthor{\binits{H.}~\bsnm{Hu}} and
\bauthor{\binits{P.}~\bsnm{Su}},
\bctitle{COOPER: Testing the Binding Code of Scripting Languages with Cooperative Mutation.},
in: \bbtitle{Proceedings of the 28th Annual Network and Distributed System Security Symposium (NDSS)},
\byear{2022}.
\end{bchapter}
\endbibitem

\bibitem{FREEDOM}
\begin{bchapter}
\bauthor{\binits{W.}~\bsnm{Xu}},
\bauthor{\binits{S.}~\bsnm{Park}} and
\bauthor{\binits{T.}~\bsnm{Kim}},
\bctitle{Freedom: Engineering a state-of-the-art dom fuzzer},
in: \bbtitle{Proceedings of the 2020 ACM SIGSAC Conference on Computer and Communications Security},
\byear{2020},
pp.~\bfpage{971}--\blpage{986}.
\end{bchapter}
\endbibitem

\bibitem{FileSystems}
\begin{bchapter}
\bauthor{\binits{W.}~\bsnm{Xu}},
\bauthor{\binits{H.}~\bsnm{Moon}},
\bauthor{\binits{S.}~\bsnm{Kashyap}},
\bauthor{\binits{P.-N.}~\bsnm{Tseng}} and
\bauthor{\binits{T.}~\bsnm{Kim}},
\bctitle{Fuzzing file systems via two-dimensional input space exploration},
in: \bbtitle{Proceedings of the 2019 IEEE Symposium on Security and Privacy (SP)},
\binstitute{IEEE},
\byear{2019},
pp.~\bfpage{818}--\blpage{834}.
\end{bchapter}
\endbibitem

\bibitem{QSYM}
\begin{bchapter}
\bauthor{\binits{I.}~\bsnm{Yun}},
\bauthor{\binits{S.}~\bsnm{Lee}},
\bauthor{\binits{M.}~\bsnm{Xu}},
\bauthor{\binits{Y.}~\bsnm{Jang}} and
\bauthor{\binits{T.}~\bsnm{Kim}},
\bctitle{QSYM: A practical concolic execution engine tailored for hybrid fuzzing},
in: \bbtitle{Proceedings of the 27th USENIX Security Symposium (USENIX Security 18)},
\byear{2018},
pp.~\bfpage{745}--\blpage{761}.
\end{bchapter}
\endbibitem

\bibitem{AFL}
\begin{botherref}
\oauthor{\binits{M.}~\bsnm{Zalewski}},
American Fuzzy Lop,
2023.
\url{https://github.com/google/AFL}.
\end{botherref}
\endbibitem

\bibitem{Squirrel}
\begin{bchapter}
\bauthor{\binits{R.}~\bsnm{Zhong}},
\bauthor{\binits{Y.}~\bsnm{Chen}},
\bauthor{\binits{H.}~\bsnm{Hu}},
\bauthor{\binits{H.}~\bsnm{Zhang}},
\bauthor{\binits{W.}~\bsnm{Lee}} and
\bauthor{\binits{D.}~\bsnm{Wu}},
\bctitle{Squirrel: Testing database management systems with language validity and coverage feedback},
in: \bbtitle{Proceedings of the 2020 ACM SIGSAC Conference on Computer and Communications Security},
\byear{2020},
pp.~\bfpage{955}--\blpage{970}.
\end{bchapter}
\endbibitem

\bibitem{Fuzzguard}
\begin{bchapter}
\bauthor{\binits{P.}~\bsnm{Zong}},
\bauthor{\binits{T.}~\bsnm{Lv}},
\bauthor{\binits{D.}~\bsnm{Wang}},
\bauthor{\binits{Z.}~\bsnm{Deng}},
\bauthor{\binits{R.}~\bsnm{Liang}} and
\bauthor{\binits{K.}~\bsnm{Chen}},
\bctitle{$\{$FuzzGuard$\}$: Filtering out unreachable inputs in directed grey-box fuzzing through deep learning},
in: \bbtitle{Proceedings of the 29th USENIX security symposium (USENIX security 20)},
\byear{2020},
pp.~\bfpage{2255}--\blpage{2269}.
\end{bchapter}
\endbibitem

\bibitem{CVE-2012-2081}
\begin{botherref}
Bug CVE-2012-2081.
\url{https://nvd.nist.gov/vuln/detail/CVE-2012-2081}.
\end{botherref}
\endbibitem

\bibitem{CVE-2014-4987}
\begin{botherref}
Bug CVE-2012-2081.
\url{https://nvd.nist.gov/vuln/detail/CVE-2014-4987}.
\end{botherref}
\endbibitem

\bibitem{SQLancer1}
\begin{botherref}
Bugs Found in Database Management Systems.
\url{https://www.manuelrigger.at/dbms-bugs/}.
\end{botherref}
\endbibitem

\bibitem{LLVMpass}
\begin{botherref}
LLVM pass.
\url{https://llvm.org/docs/WritingAnLLVMPass.html}.
\end{botherref}
\endbibitem

\bibitem{95908}
\begin{botherref}
MySQL bug 95908.
\url{https://bugs.mysql.com/bug.php?id=95908}.
\end{botherref}
\endbibitem

\bibitem{95926}
\begin{botherref}
MySQL bug 95926.
\url{https://bugs.mysql.com/bug.php?id=95926}.
\end{botherref}
\endbibitem

\bibitem{95927}
\begin{botherref}
MySQL bug 95927.
\url{https://bugs.mysql.com/bug.php?id=95927}.
\end{botherref}
\endbibitem

\bibitem{95937}
\begin{botherref}
MySQL bug 95937.
\url{https://bugs.mysql.com/bug.php?id=95937}.
\end{botherref}
\endbibitem

\bibitem{95954}
\begin{botherref}
MySQL bug 95954.
\url{https://bugs.mysql.com/bug.php?id=95954}.
\end{botherref}
\endbibitem

\bibitem{95975}
\begin{botherref}
MySQL bug 95975.
\url{https://bugs.mysql.com/bug.php?id=95975}.
\end{botherref}
\endbibitem

\bibitem{95983}
\begin{botherref}
MySQL bug 95983.
\url{https://bugs.mysql.com/bug.php?id=95983}.
\end{botherref}
\endbibitem

\bibitem{96012}
\begin{botherref}
MySQL bug 96012.
\url{https://bugs.mysql.com/bug.php?id=96012}.
\end{botherref}
\endbibitem

\bibitem{99122}
\begin{botherref}
MySQL bug 99122.
\url{https://bugs.mysql.com/bug.php?id=99122}.
\end{botherref}
\endbibitem

\bibitem{SQLancer}
\begin{botherref}
SQLancer.
\url{https://github.com/sqlancer/sqlancer}.
\end{botherref}
\endbibitem

\bibitem{16252d7}
\begin{botherref}
SQLite bug 16252d7.
\url{https://www.sqlite.org/src/info/16252d7}.
\end{botherref}
\endbibitem

\bibitem{1685610e}
\begin{botherref}
SQLite bug 1685610e.
\url{https://www.sqlite.org/src/info/1685610e}.
\end{botherref}
\endbibitem

\bibitem{1b1dd4d4}
\begin{botherref}
SQLite bug 1b1dd4d4.
\url{https://www.sqlite.org/src/info/1b1dd4d4}.
\end{botherref}
\endbibitem

\bibitem{2363a14}
\begin{botherref}
SQLite bug 2363a14.
\url{https://www.sqlite.org/src/info/2363a14}.
\end{botherref}
\endbibitem

\bibitem{5351e920}
\begin{botherref}
SQLite bug 5351e920.
\url{https://www.sqlite.org/src/info/5351e920}.
\end{botherref}
\endbibitem

\bibitem{54110870}
\begin{botherref}
SQLite bug 54110870.
\url{https://www.sqlite.org/src/info/54110870}.
\end{botherref}
\endbibitem

\bibitem{5c6146b5}
\begin{botherref}
SQLite bug 5c6146b5.
\url{https://www.sqlite.org/src/info/5c6146b5}.
\end{botherref}
\endbibitem

\bibitem{659c551d}
\begin{botherref}
SQLite bug 659c551d.
\url{https://www.sqlite.org/src/info/659c551d}.
\end{botherref}
\endbibitem

\bibitem{6ac0f822}
\begin{botherref}
SQLite bug 6ac0f822.
\url{https://www.sqlite.org/src/info/6ac0f822}.
\end{botherref}
\endbibitem

\bibitem{86fa0087}
\begin{botherref}
SQLite bug 86fa0087.
\url{https://www.sqlite.org/src/info/86fa0087}.
\end{botherref}
\endbibitem

\bibitem{9c8c1092}
\begin{botherref}
SQLite bug 9c8c1092.
\url{https://www.sqlite.org/src/info/9c8c1092}.
\end{botherref}
\endbibitem

\bibitem{c0c90961}
\begin{botherref}
SQLite bug c0c90961.
\url{https://www.sqlite.org/src/info/c0c90961}.
\end{botherref}
\endbibitem

\bibitem{db9acef1}
\begin{botherref}
SQLite bug db9acef1.
\url{https://www.sqlite.org/src/info/db9acef1}.
\end{botherref}
\endbibitem

\bibitem{eb40248}
\begin{botherref}
SQLite bug eb40248.
\url{https://www.sqlite.org/src/info/eb40248}.
\end{botherref}
\endbibitem

\bibitem{ebe4845c}
\begin{botherref}
SQLite bug ebe4845c.
\url{https://www.sqlite.org/src/info/ebe4845c}.
\end{botherref}
\endbibitem

\bibitem{f65c929}
\begin{botherref}
SQLite bug f65c929.
\url{https://www.sqlite.org/src/info/f65c929}.
\end{botherref}
\endbibitem

\bibitem{f898d04c}
\begin{botherref}
SQLite bug f898d04c.
\url{https://www.sqlite.org/src/info/f898d04c}.
\end{botherref}
\endbibitem

\bibitem{f9c6426}
\begin{botherref}
SQLite bug f9c6426.
\url{https://www.sqlite.org/src/info/f9c6426}.
\end{botherref}
\endbibitem

\bibitem{faaaae49}
\begin{botherref}
SQLite bug faaaae49.
\url{https://www.sqlite.org/src/info/faaaae49}.
\end{botherref}
\endbibitem

\bibitem{15725}
\begin{botherref}
TiDB bug 15725.
\url{https://github.com/pingcap/tidb/issues/15725}.
\end{botherref}
\endbibitem

\bibitem{15733}
\begin{botherref}
TiDB bug 15733.
\url{https://github.com/pingcap/tidb/issues/15733}.
\end{botherref}
\endbibitem

\bibitem{15789}
\begin{botherref}
TiDB bug 15789.
\url{https://github.com/pingcap/tidb/issues/15789}.
\end{botherref}
\endbibitem

\bibitem{15846}
\begin{botherref}
TiDB bug 15846.
\url{https://github.com/pingcap/tidb/issues/15846}.
\end{botherref}
\endbibitem

\bibitem{15986}
\begin{botherref}
TiDB bug 15986.
\url{https://github.com/pingcap/tidb/issues/15986}.
\end{botherref}
\endbibitem

\bibitem{15994}
\begin{botherref}
TiDB bug 15994.
\url{https://github.com/pingcap/tidb/issues/15994}.
\end{botherref}
\endbibitem

\bibitem{17814}
\begin{botherref}
TiDB bug 17814.
\url{https://github.com/pingcap/tidb/issues/17814}.
\end{botherref}
\endbibitem

\end{thebibliography}

%

\end{document}